\documentclass[lettersize,journal]{IEEEtran}
\usepackage{amsmath,amsfonts}
\usepackage{algorithmic}
\usepackage{algorithm}
\usepackage{array}
\usepackage{cuted}
\usepackage{subfigure}
\usepackage{textcomp}
\usepackage{stfloats}
\usepackage{url}
\usepackage{verbatim}
\usepackage{graphicx}
\usepackage{cite}
\usepackage{bm}
\usepackage{epstopdf}
\usepackage{float}
\usepackage{booktabs}
\usepackage{multirow}
\newcommand{\qa}{{\bf a}}

\newcommand{\qF}{{\bf F}}
\newcommand{\qI}{{\bf I}}
\newcommand{\qg}{{\bf g}}
\newcommand{\qh}{{\bf h}}
\newcommand{\qH}{{\bf H}}
\newcommand{\ql}{{\bf l}}

\newcommand{\qs}{{\bf s}}
\newcommand{\qz}{{\bf z}}
\newcommand{\qT}{{\rm \top}}
\newcommand{\qv}{{\bm v}}

\newcommand{\qy}{{\bm y}}
\newcommand{\qY}{{\bf Y}}

\usepackage{xcolor}
\usepackage{color}
\newtheorem{theorem}{\textbf{Theorem}}

\begin{document}

\title{RIS-Empowered Integrated Location Sensing and Communication with Superimposed Pilots}
\author{Wenchao Xia,~\IEEEmembership{Member,~IEEE,}
Ben Zhao,
Wankai Tang,~\IEEEmembership{Member,~IEEE,}
Yongxu Zhu,~\IEEEmembership{Senior Member,~IEEE}
Kai-Kit Wong,~\IEEEmembership{Fellow,~IEEE}
Sangarapillai Lambotharan,~\IEEEmembership{Senior Member,~IEEE}
Hyundong Shin,~\IEEEmembership{Fellow,~IEEE}

\thanks{W. Xia and B. Zhao are with the Jiangsu Key Laboratory of Wireless Communications and IoT, and also with the Engineering Research Center of Health Service System Based on Ubiquitous Wireless Networks, Ministry of Education,  Nanjing University of Posts and Telecommunications, Nanjing 210003, China. (e-mail: xiawenchao@njupt.edu.cn, 1023010232@njupt.edu.cn)}
\thanks{W. Tang and Y. Zhu are with the National Mobile Communications Research Laboratory, Southeast University, Nanjing 210096, China. (e-mail: tangwk@seu.edu.cn, yongxu.zhu@seu.edu.cn)}
\thanks{K.-K. Wong is with the Department of Electronic and Electrical Engineering, University College London, WC1E 7JE London, U.K., and also with the Department of Electronic Engineering, Kyung Hee University, Yongin-si, Gyeonggi-do 17104, Republic of Korea.  (e-mail: kai-kit.wong@ucl.ac.uk)}
\thanks{S. Lambotharan is with the Institute for Digital Technologies, Loughborough University, London, E20 3BS, U.K. (e-mail: S.Lambotharan@lboro.ac.uk)}
\thanks{H.~Shin  is with the Department of Electronics and Information Convergence Engineering,  Kyung Hee University,
1732 Deogyeong-daero, Giheung-gu, Yongin-si, Gyeonggi-do 17104, Republic of Korea (e-mail: hshin@khu.ac.kr). }

\vspace{-20pt}
}



\maketitle

\begin{abstract}
In addition to enhancing wireless communication coverage quality, reconfigurable intelligent surface (RIS) technique can also assist in positioning. In this work, we consider RIS-assisted superimposed pilot and data transmission without the assumption availability of prior channel state information and position information of mobile user equipments (UEs). To tackle this challenge, we design a frame structure of transmission protocol composed of several
location coherence intervals, each with pure-pilot and data-pilot transmission durations. The former is used to estimate UE locations, while the latter is time-slotted, duration of which does not exceed the channel coherence time, where the data and pilot signals are transmitted simultaneously.  We conduct the Fisher Information matrix (FIM) analysis and derive  \text {Cramér-Rao bound} (CRB) for the position estimation error. The inverse fast Fourier transform (IFFT) is adopted to obtain the estimation results of UE positions, which are then exploited for channel estimation. Furthermore, we derive the closed-form lower bound of the ergodic achievable rate of superimposed pilot (SP) transmission, which is used to optimize the phase profile of the RIS to maximize the achievable sum rate using the genetic algorithm. Finally, numerical results validate the accuracy of the UE position estimation using the IFFT algorithm and the superiority of the proposed SP  scheme by comparison with the regular pilot scheme.
\end{abstract}

\begin{IEEEkeywords}
Reconfigurable intelligent surface (RIS), Integrated sensing and communication (ISAC), dual-function radar and communication (DFRC), superimposed pilot, positioning
\end{IEEEkeywords}

\section{Introduction}

Reconfigurable intelligent surface (RIS) is a promising technique for B5G and 6G  because of its ability to customize wireless channels such that it can passively steer electromagnetic waves in desired directions of space.  By placing  RIS in an appropriate location between a base station (BS) and user equipments (UEs),  one can establish virtual Line-of-sight (LOS) links for the BS and UEs, so as to enhance the enlargement of wireless coverage and the increase of network capacity, reduce transmit power, and improve transmission reliability. 

RIS intended for communication purposes has been widely studied \cite{tang2021wireless,bjornson2022reconfigurable,li2024ergodic}.  However, many existing works assumed availability of instantaneous CSI, which should be estimated in each channel coherence interval. CSI estimation is not a trivial task for at least the two following reasons. The first challenge is communication overhead especially in scenarios with massive UEs. In a regular pilot (RP) transmission scheme \cite{wang2020channel,hu2021two,you2020channel}, the pilot and data symbols are transmitted separately to mitigate interference, which leads to low resource efficiency. To improve resource efficiency,   a novel transmission scheme, namely superimposed pilot (SP), is proposed \cite{hoeher1999channel,tong2004pilot}, which simultaneously transmits pilot and data signals, such that more samples can be used for channel estimation while maintaining a good data transmission. The work in \cite{asyhari2017orthogonal}  provided a detailed comparison of SP and RP in terms of achievable rate and normalized mean-squared error with respect to transmit antenna number and fading variation.  In \cite{verenzuela2018spectral}, the authors derived rigorous closed-form rate expressions with SP and evaluated the performance of SP in the uplink of a multicell massive multiple-input multiple-out (MIMO) system. The authors of \cite{jing2018superimposed} proposed a block-diagonal Grassmannian line packing approach to form SP matrix and then further proposed an iterative channel estimation method based on Tikhonov regularization. The SP scheme was further applied to cell-free massive MIMO systems \cite{zhang2021superimposed}. However, as far as we know, there is rarely work related to channel estimation and performance analysis in RIS-assisted SP transmission scenarios.

On the other hand, the second challenge arises from the absence of prior information. In the aforementioned works,  UE locations are assumed to be fixed and known a priori. However, this assumption is not always valid because some UEs, such as vehicles and drones, are mobile. To design high-rate directional beamforming,  prior knowledge of UE locations should be estimated.  Conventional received signal strength-based localization methods are heavily dependent on propagation conditions and thus suffer from poor accuracy \cite{weiss2003accuracy}. According to 3GPP standards, time-difference-of-arrival measurements can be used for UE positioning, but at least four BSs are required for three-dimensional (3D) UE location if a UE is not well-synchronized to these BSs \cite{keykhosravi2021siso}. In 5G, millimeter wave (mmWave) based localization with a single BS is possible given that BS and UE have antenna arrays and rich multipath condition is assumed. Angle-of-arrival-based localization methods can achieve higher accuracy, but these methods excessively depend on LoS links, which may be blocked and unavailable in practical scenarios \cite{hu2023irs}. Thankfully, RIS technique offers promising solutions.

RIS is considered to be one of the enabling techniques for UE positioning \cite{hu2018beyond,wymeersch2020beyond,he2020large,chen2024joint,fascista2022ris,jiang2022optimization,peng2024joint}. In \cite{hu2018beyond}, Fisher information matrix (FIM) and \text {Cramér-Rao bound} (CRB) analysis was conducted for UE positioning.  In \cite{wymeersch2020beyond,he2020large}, an RIS with a uniform linear array was used to estimate one UE location. In addition, RIS combination and phase control were jointly optimized to improve positioning performance. 
The work  \cite{chen2024joint} exploited the partial overlapping structured sparsity and two-dimensional (2D) block sparsity inherent in the sensing and communication channels and an angle-based sensing turbo variational Bayesian inference was developed to solve the considered joint location sensing and channel estimation problem.
A joint localization and synchronization based on an optimized design of the BS active precoding and RIS passive phase profile was proposed in \cite{fascista2022ris}.   The authors in \cite{jiang2022optimization} advanced it  by considering the CSI acquisition.
The authors in \cite{peng2024joint} proposed an algorithm for joint location sensing and demodulation for IRS-assisted ISAC mmWave MIMO systems, but two RISs were exploited and only one UE equipped with multiple antennas was considered.

Most existing works on RIS-assisted UE positioning focus on analyzing statistical position error (such as CRB) but do not give explicit estimation results of UE positions, which can be used for channel estimation and further data decoding. Besides,  the RIS-aided SP transmission is not well investigated and explored.  Motivated by the above mentioned issues, in this work we consider an uplink multiple-input single-output (MISO) system combined with an RIS and  provide a location-sensing empowered SP transmission scheme.   The main contributions are summarised as follows.
\begin{itemize}
    \item  Considering the mobility of UEs, we design a frame structure of RIS-assisted SP transmission protocol composed of several location coherence intervals each with pure-pilot and data-pilot transmission durations.  The pure-pilot transmission duration is used to estimate UE locations, while the data-pilot transmission duration is time-slotted duration of which does not exceed the  channel coherence time  and the data and pilot signals are transmitted simultaneously. 
    \item  We conduct the FIM analysis and derive the CRB of position estimation error. In addition, the inverse fast Fourier transform (IFFT) algorithm is adopted to obtain the estimation results of UE positions, which are then exploited for the channel estimation.
    \item  We derive the closed-form lower bound of the ergodic achievable sum rate of the RIS-assisted uplink SP transmission system, based on which we optimize the phase profile of the RIS using the genetic algorithm (GA).
    \item Finally, numerical results are presented to validate the accuracy of the location estimation results using the IFFT algorithm and the performance superiority of the proposed SP scheme compared with the RP  scheme.
\end{itemize}

The remainder of this work is organized as follows. Section \ref{section:system model} introduces the considered system model and transmission protocol. Section \ref{section: location sensing} presents the location sensing method and analyzes the estimation error.  The closed-form lower bound of the ergodic achievable rate of RIS-aided SP transmission is derived and the optimization method of the RIS phase profile is proposed in Section \ref{section: SP}.   Besides, simulation results are conducted to validate the efficiency of the proposed methods in Section \ref{section: simulations}. Finally, Section \ref{section: conclusions} concludes this work.

\textbf{Notations:} Matrices and
vectors are denoted by bold capital and lowercase symbols, respectively.  $(\mathbf{A})^{\rm T}$ and $(\mathbf{A})^{\rm H}$  stand for transpose and conjugate transpose of $\mathbf{A}$, respectively.  The notations $\otimes$, $\odot$, and $||\bullet||$ are the Kronecker product, Hadamard  product, and $l_2$-norm operators, respectively. The operator $\text{diag}(\qa)$ denotes the operation to diagonalize the vector $\qa$ into a matrix whose main diagonal elements are from $\mathbf{a}$.   Finally, we denote a complex Gaussian vector by $\qa\sim\mathcal{CN}(\bar{\qa},\bm{\Sigma})$  with the mean vector $\bar{\qa}$ and covariance matrix $\bm{\Sigma}$; if $\qa$ is a real vector, $\qa\sim\mathcal{N}(\bar{\qa},\bm{\Sigma})$ is used.

\section{System Model}\label{section:system model}
\begin{figure}[!t]
\centering
\includegraphics[width=0.47\textwidth]{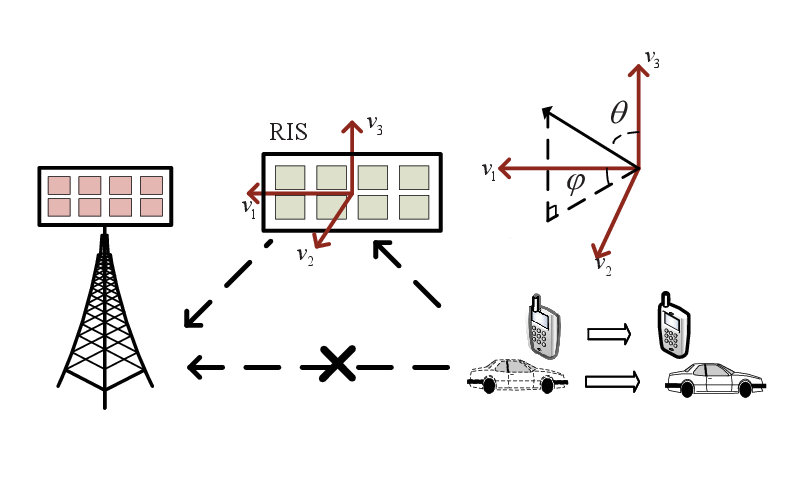}
\caption{A snapshot of the considered system model.}
\label{system_model}
\end{figure}
\begin{figure}[!t]
\centering
\includegraphics[width=0.47\textwidth]{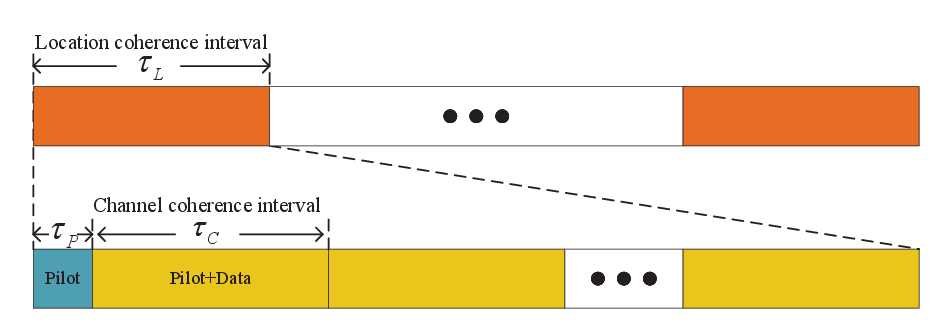}
\caption{The frame structure of the considered transmission protocol.}
\label{frame_structure}
\end{figure}
\subsection{System Setup}
As shown in Fig. \ref{system_model}, we consider a narrowband system consisting of one BS, one RIS, and $K$ single-antenna devices, indexed by $\mathcal{K}=\{1,2,\ldots, K\}$. It is assumed that the LOS path from the BS to the UEs is blocked, but the LOS path from the BS to the RIS and the LOS path from the RIS to the UEs are available, which conduct a virtual LOS  path for the BS-UE links. In addition, we assume that the BS, RIS, and UEs are in the far-field region and treat the transmitted beam as a uniform plane wave, ignoring the near-field radiation model of electromagnetic waves \cite{9903389}. In particular, The BS has a uniform planar array (UPA) with  $M_{\rm B}=M_{{\rm B},{\rm x}}\times M_{{\rm B},{\rm z}} $ antenna elements. The RIS is also equipped with a UPA with $M_{\rm R}=M_{{\rm R},{\rm x}}\times M_{{\rm R},{\rm z}} $ reflecting elements. The element spacing for the UPAs of the BS and RIS is set as $d=\lambda_{\rm c}/2$  where $\lambda_{\rm c}$ is the carrier wavelength. 

Denote by $\ql_{\rm B}\in \mathbb{R}^{3\times 1}$, $\ql_{\rm R}\in \mathbb{R}^{3\times 1}$ and $\ql_{\it k}\in \mathbb{R}^{3\times 1}$ the locations of the array center of the BS, the array center of the RIS, and UE $k$ in the global Cartesian coordinate system, respectively. We also denote by ${\bf V}_{\rm B}=[\qv_{\rm B1},\qv_{\rm B2},\qv_{\rm B3}]^\qT$ and ${\bf V}_{\rm R}=[\qv_{\rm R3},\qv_{\rm R3},\qv_{\rm R3}]^\qT$  the rotation matrices corresponding to the BS and RIS coordinate systems, respectively. The BS and RIS are fixed and their locations are known a priori. However, these UEs are mobile, whose locations are unknown and need to be estimated before the BS provides service for these UEs. Furthermore, to improve spectral efficiency,  the data and pilot are transmitted simultaneously, where the pilots are used for channel estimation with the sensed UE positions.

\subsection{Protocol Design}
In this work, we assume that the UE locations are fixed or change slowly within one location coherence interval, but vary obviously when crossing different location coherence intervals. 
Similarly, the channel remains unchanged within one channel coherence interval but is independent in different channel coherence intervals.
As shown in Fig. \ref{frame_structure}, a transmission protocol is designed, where the implementation of communication and perception functions is time-division. We assume that each location coherence interval $\tau_{\rm L}$ is composed of one sensing interval with the duration of ${\tau}_{\rm P}$ and $N_{\rm C}$ channel coherence intervals each with the duration of ${\tau}_{\rm C}$, i.e., ${\tau}_{\rm L}={\tau}_{\rm P}+N_{\rm C}{\tau}_{\rm C}$. Thus, the number of transmissions within one location coherence interval is $\tau_{\rm L} B$, where $B$ is the bandwidth.
The sensing interval is for UE localization and the remaining is for uplink data transmission with SPs.   Note that each channel coherence interval performs one transmission operation where data and pilots are sent simultaneously.  Channel estimation is based on superimposed pilots and then data detection relies on the estimated channel results.  

Here, the movement of each UE is modelled as a random walk process.  Denoted by ${{{\bf l}_k}({0})}$ the position of the $k$-th UE at the beginning of a frame, then the position in the $n_{\rm L}$-th location coherence interval is updated as ${{{\bf l}_k}({n_{\rm L}})}={{{\bf l}_k}({n_{\rm L}}-1)}+{\Delta}{{{\bf l}_k}({n_{\rm L}})}$ with ${\Delta}{\bf l}_k({n_{\rm L}})\sim{\mathcal{N}}(0,{\bf \Xi}_{k})$ and ${\bf \Xi}_{k}$ being the covariance of the random walk in the period of $\tau_{\rm L}$.

\subsection{Channel Model}
For a UPA with $M=M_{\rm x}\times M_{\rm z}$ elements, we define ${\bf n}_{m_{\rm x},m_{\rm z}}=[(m_{\rm x}-1)d-(M_{\rm x}-1)d/2,0,(m_{\rm z}-1)d-(M_{\rm z}-1)d/2]^\qT$ as the position of the $(m_{\rm x},m_{\rm z})$-th element in the UPA's local coordinate system and collect these positions into a set ${\bf N}=[{\bf n}_{1,1},{\bf n}_{1,2},\ldots,{\bf n}_{M_{\rm x},M_{\rm z}}]^\qT\in\mathbb{R}^{(M_{\rm x}M_{\rm z})\times 3}$ in a Cartesian coordinate system.  
We also define a vector $\bm{\psi}=[\varphi,\theta]^\qT$ in a spherical coordinate system with $\varphi$ and $\theta$ being the azimuth and elevation angles, respectively. Given $\bf N$ and $\bm{\psi}$, we further define the steering vector as ${\bm a}(\bm{\psi})=[{\it a_{\rm 1}},{\it a_{\rm 2}},\ldots,{\it a}_{\it M_{\rm x}\it M_{\rm z}}]^\qT$, where the $((m_{\rm x}-1)M_{\rm z}+m_{\rm z})$-th element is 
\begin{equation}
a_{(m_{\rm x}-1){\times}M_{\rm z}+m_{\rm z}}=e^{j\frac{2\pi}{\lambda_{\rm c}}{\bf n}_{m_{\rm x},m_{\rm z}}^\qT{{\bm \omega}({\bm \psi})}},
\end{equation}
in which 
\begin{equation}\label{omega_vec}
{\bm \omega}({\bm \psi})=[\sin(\theta)\cos(\varphi),\sin(\theta)\sin(\varphi),\cos(\theta)]^\qT.
\end{equation} 
Then, the response vector  for the angle ${\bm \psi}$ can be expressed as ${\bm a}(\bm \psi)={{\bm a}_{\rm x}(\bm \psi)}\otimes{{\bm a}_{\rm z}(\bm \psi)}$ with 
\begin{equation}
{\bm a}_{\rm x}({\bm \psi})=e^{j\zeta_{\rm x}}[1,e^{-j[{\bm \mu}({\bm \psi})]_1d},\ldots,e^{-j[{\bm \mu}({\bm \psi})]_1(M_{\rm x}-1)d}]^\qT,
\end{equation}
\begin{equation}
{\bm a}_{\rm z}({\bm \psi})=e^{j\zeta_{\rm z}}[1,e^{-j[{\bm \mu}({\bm \psi})]_3d},\ldots,e^{-j[{\bm \mu}({\bm \psi})]_3(M_{\rm z}-1)d}]^\qT,
\end{equation}
where ${\bm \mu}({\bm \psi})=-\frac{2\pi}{\lambda_{\rm c}}{\bm \omega}({\bm \psi})$, $\zeta_{\rm x}=[{\bm \mu}(\bm \psi)]_1(M_{\rm x}-1)d/2$, and $\zeta_{\rm z}=[{\bm \mu}(\bm \psi)]_3(M_{\rm z}-1)d/2$.

Based on the above definitions, the downlink channel between the BS and the RIS, modelled as the Rician channel, is represented as
\begin{equation}
{\bf H}_0={\bar{\bf H}_0}+{\tilde{\bf H}_0},
\end{equation}
where $\bar{\qH}_0\in\mathbb{C}^{M_{\rm R}\times M_{\rm B}}$ and $\tilde{\qH}_0\in\mathbb{C}^{M_{\rm R}\times M_{\rm B}}$ are the LOS and NLOS components, respectively. $\bar{\qH}_0$ is given by
\begin{equation}
{{\bar{\bf H}}_{0}}={\beta_0}{\bm a}({\bm \psi}_{\rm R}){({\bm a}({\bm \psi}_{\rm {B}}))}^{\qT},
\end{equation}
where ${\beta_0}={\alpha_0}{\sqrt{{\varepsilon_0}/{(\varepsilon_0+1)}}}$ represents the channel gain of the BS-RIS LOS path, $\alpha_0={\lambda}/(4{\pi}|{{\bf l}_{\rm R}-{\bf l}_{\rm B}}|)\exp({-j2\pi{|{{\bf l}_{\rm R}-{\bf l}_{\rm B}}|}/{\lambda}})$ is the path gain, ${{\varepsilon}_0}({\varepsilon_0}>0)$ is the Rice factor,  ${\bm \psi}_{\rm B}=[{{\varphi}_{\rm B}},{{\theta}_{\rm B}}]^\qT$ as the angle-of-departure (AOD) from the BS to the RIS in the local coordinate system of the BS, and ${\bm \psi}_{\rm R}=[{{\varphi}_{\rm R}},{{\theta}_{\rm R}}]^\qT$ as the angle-of-arrival (AOA) at the RIS from the BS in the local coordinate system of the RIS. The NLOS component ${\tilde{\bf H}_{0}}$ is modelled as a random matrix, whose elements are independently and identically distributed (i.i.d) $\mathcal{CN}(0,{\alpha_0^2}/({\varepsilon_0+1}))$ random variables. Note that the vector ${\bm \psi}_{\rm B}$ corresponds to the direction of  $\Delta{\bf l}_{\rm B}={\bf V}_{\rm B}({\bf l}_{\rm R}-{\bf l}_{\rm B})$ in the local Cartesian  coordinate system of the BS. We define ${\rho}_{\rm B}(r_{\rm B}, \varphi_{\rm B}, \theta_{\rm B})$ as the local spherical coordinate system of BS, where the radial distance $r_{\rm B}$, the elevation angle ${\varphi}_{\rm B}$, and the azimuth angle ${\theta}_{\rm B}$ are given by ${r_{\rm B}}=\|\Delta{\bf l}_{\rm B}\|$, 
\begin{equation}
{\varphi}_{\rm B}=\arctan2([\Delta{\bf l}_{\rm B}]_2,[\Delta{\bf l}_{\rm B}]_1),
\end{equation}
and
\begin{equation}
{\theta}_{\rm B}=\arccos([\Delta{\bf l}_{\rm B}]_3/\|\Delta{\bf l}_{\rm B}\|),
\end{equation}
respectively. In addition, we define ${\rho}_{\rm R}(r_{\rm R}, \varphi_{\rm R}, \theta_{\rm R})$ as the local spherical coordinate system of the RIS, where ${r_{\rm R}}=\|\Delta{\bf l}_{\rm R}\|$, and the azimuth angle ${\varphi}_{\rm R}$ and elevation angle ${\theta}_{\rm R}$ of the AOA at the RIS from the BS are given by 
\begin{equation}
{\varphi}_{\rm R}=\arctan2([\Delta{\bf l}_{\rm R}]_2,[\Delta{\bf l}_{\rm R}]_1),
\end{equation}
 and 
\begin{equation}
{\theta}_{\rm R}=\arccos([\Delta{\bf l}_{\rm R}]_3/\|\Delta{\bf l}_{\rm R}\|),
\end{equation}
respectively, where $\Delta{\bf l}_{\rm R}={\bf V}_{\rm R}({\bf l}_{\rm B}-{\bf l}_{\rm R})$.
Then, we have
\begin{equation}\label{xlabel}
  {{\bf l}_{\rm R}}={\bf l}_{\rm B}+{\bf V}^{\qT}_{\rm B}{r_{\rm B}}
  \left [ \begin{matrix}
  \sin(\theta_{\rm B})\cos(\varphi_{\rm B}) \\
  \sin(\theta_{\rm B})\sin(\varphi_{\rm B}) \\
  \cos(\theta_{\rm B}) \\
  \end{matrix} \right ].
\end{equation} 

Similarly, the channel between the RIS and the $k$-UE is also modelled as the Rician channel, i.e.,
\begin{equation}
\qh_k={{\bar{\qh}_k}+{\tilde{\qh}_k}},\forall{k},
\end{equation}
in which  ${\tilde{\qh}_k}\in\mathbb{C}^{M_{\rm R}\times 1}$  is the NLoS component whose elements each follow $\mathcal{CN}(0,{{\alpha_k^2}/{(\varepsilon_k+1)}})$ and $ {\bar{\qh}_k}$ is the LOS component given by
\begin{equation}
 {\bar{\qh}_k}={\beta_k}{{\bm a}(\bm{\psi}_{\it k})},
\end{equation}
where ${\beta_k}={\alpha_k}{\sqrt{{\varepsilon_k}/{(\varepsilon_k+1)}}}$ represents the channel gain of  LOS path between the RIS and UE $k$, $\alpha_k={\lambda}/(4{\pi}\|{{\bf l}_{\it k}-{\bf l}_{\rm R}}\|^{b/2})\exp({-j2\pi{|{{\bf l}_{\it k}-{\bf l}_{\rm R}}|}/{\lambda}})$, where $b$ is the path-loss exponent and $\varepsilon_k({\varepsilon_k>0})$ is the Rician factor. We define ${\rho}_{k}(r_{k}, \varphi_{k}, \theta_{k})$ as the local spherical coordinate system of UE $k$, where ${r_{k}}=\|\Delta{\bf l}_{k}\|$,  and ${\bm \psi}_{\it k}=[{{\varphi}_{\it k}},{{\theta}_{\it k}}]^\qT$ is the AOD from the RIS to the $k$-th UE, whose azimuth and elevation angles are given by
\begin{equation}
{{\varphi}_{\it k}}=\arctan2([\Delta{\bf l}_{k}]_2,[\Delta{\bf l}_{k}]_1),
\end{equation}
and
\begin{equation}
{{\theta}_{\it k}}=\arccos([\Delta{\bf l}_{k}]_3/\|\Delta{\bf l}_{k}\|),
\end{equation}
respectively, where $\Delta{\bf l}_{k}={\bf V}_{\rm R}({\bf l}_{k}-{\bf l}_{\rm R})$.
Then, we have
\begin{equation}\label{xlabel}
  {{\bf l}_{k}}={\bf l}_{\rm R}+{\bf V}^{\qT}_{\rm R}{r_{\rm R}}
  \left [ \begin{matrix}
  \sin(\theta_{\rm R})\cos(\varphi_{\rm R}) \\
  \sin(\theta_{\rm R})\sin(\varphi_{\rm R}) \\
  \cos(\theta_{\rm R}) \\
  \end{matrix} \right ].
\end{equation} 

The whole channel vector from the BS to the $k$-th UE  is represented as\footnote{Note that we do not consider hardware impairments\cite{li2023achievable,10025392} in  this work due to high complexity of deriving closed-form lower bound of the ergodic achievable sum rate. }
\begin{equation}
\qh_{{\rm B},{\it k}}=
{{\bf H}^{\rm H}_{0}}{{\rm {diag}}({\rm {vec}}({\bf \Lambda}))}{\qh_{k}},
\end{equation}
where $\qh_{{\rm B},{\it k}}\in\mathbb{C}^{ M_{\rm B} \times 1}$ and ${\bf\Lambda}\in\mathbb{C}^{M_{\rm {R,x}} \times M_{\rm {R,z}}}$ is the RIS phase profile with $|[{\bf\Lambda}]_{m_{\rm x},m_{\rm z}}|=1,\forall m_{\rm x},m_{\rm z}$.

Here, we assume that the two parts of the cascaded channel are independent of each other and have $\qh_{\rm B,\it k}\sim{\mathcal{CN}}\left ({\bar{\qh}_{{\rm B},k}},{{\bf R}_{\rm B,\it k}}\right)$, in which 
\begin{equation}
{{\bar{\qh}}_{\rm B,\it k}}={\mathbb{E}}\{{\qh_{\rm B,\it k}}\}={\bar{\bf H}}^{\rm H}_{\rm {0}}{\rm {diag}}({\rm {vec}}({\bf \Lambda})){{\bar{\qh}}_k},
\end{equation}
and  the variance ${{\bf R}_{\rm B,\it k}}$ is computed as
\begin{equation}
\begin{split}
{\bf R}_{\rm B,\it k}&={\mathbb{E}} \{{\left ({\qh_{\rm B,\it k}}-{{\bar{\qh}}_{{\rm B},k}}\right )\left ({\qh_{{\rm B},k}}-{{\bar{\qh}}_{{\rm B},k}}\right ) ^{\rm H}}\}.\\
\end{split}
\end{equation}


In what follows, we first give the location sensing scheme, then introduce the SP-based transmission scheme with the estimated results of UE locations.

\section{Location Sensing}\label{section: location sensing}
In this section, we first analyze the FIM of UE location, and then introduce the location estimation method in detail.

\subsection{FIM Analysis}


Without losing generality, we consider only one location coherence interval and assume that the BS transmit the same pilot sequence at each time moment. 
In the phase of location estimation, ${\tau_{\rm P}}{{\rm B}}$ pilot symbols are transmitted by the BS and we collect these pilot symbols into a vector $\qs=[{s}_1,{s}_2,\ldots,{s}_{M_{\rm B}}]^\qT$, where $|{ s}_i|^2=P_{\rm B}/({M_{\rm B}})$ and $P_{\rm B}$ is the transmit power during positioning stage. 
Then, the received signal at the $k$-th UE is given as
\begin{equation}\label{received sensing data}
y_{k,t}={\bar{\bf h}}^{\rm H}_{\rm B,\it k}\qs+w_{k,t},t\in\{1,2,\ldots,\tau_{\rm P}B\},
\end{equation}
where  $w_{k,t}\sim \mathcal{CN}(0,\sigma^2_{\rm B,\it k})$  represents the total noise variance with
\begin{equation}
\sigma^2_{\rm B,\it k}=\sigma^2_{\rm N}+\tilde{\sigma}^2_{k}+\hat{\sigma}^2_{k}+\check{\sigma}^2_{k},
\end{equation}
where $\sigma^2_{\rm N}$ represents the variance of additive white Gaussian noise (AWGN), $\tilde{\sigma}^2_{k}=||{{\bf s}}||^2{{\rho}_{{\rm B},k}}{\varepsilon_k}$ represents the noise variance from the combination  of the BS-RIS NLOS path and the RIS-UE LOS path with
\begin{equation}
\rho_{{\rm B},k}=\frac{\alpha_{k}^{2} \alpha_{0}^{2}}{(\varepsilon_{k}+1)(\varepsilon_{0}+1)},
\end{equation}
$\hat{\sigma}^2_{k}=|| {\bm a}({\bm \psi}_{\rm R})({\bm a}({\bm \psi}_{\rm B}))^{\qT}\qs||^2{\rho_{{\rm B},k}}{\varepsilon_0}$ represents the noise variance from the combination of the BS-RIS LOS path  and the RIS-UE NLOS path, and $\check{\sigma}^2_{k}=||\qs||^2\rho_{{\rm B},k}$ represents the noise variance from the combination of the BS-RIS NLOS path and the RIS-UE NLOS path.

We aim to estimate the locations of all UEs according to the observation $\qY=[\qy_1,\qy_2,\ldots,\qy_K]$ with $\qy_k=[y_{k,1},\ldots,y_{k,\tau_{\rm P}B}]^\qT$. Define $\bm{\xi}^{\rm lo}_k=[[\ql_k]_1, [\ql_k]_2,\bm{\beta}_k^\qT]^\qT$ and $\bm{\xi}^{\rm ch}_k=[\bm{\psi}_k^\qT,\bm{\beta}_k^\qT]^\qT$ as the location parameter vector and the channel parameter vector, respectively, where $\bm{\beta}_k=[\mathfrak{R}(\beta_k),\mathfrak{I}(\beta_k)]^\qT$ includes the real and imaginary parts of $\beta_k$. The FIM $\qF^{\rm ch}_k$ of $\bm{\xi}^{\rm ch}_k$ is calculated as
\begin{equation}\label{FIM}
  \qF^{\rm ch}_k=\frac{2}{\sigma_{{\rm B},{\it k}}^2}\sum^{\tau_{\rm P} B}_{t=1}\mathfrak{R}\left \{\nabla_{\bm{\xi}^{\rm ch}_k}\bar{y}_{k,t}(\nabla_{\bm{\xi}^{\rm ch}_k}\bar{y}_{k,t})^{\rm H}\right \},
\end{equation}
in which $\bar{y}_{k,t}$ is the noise-free observation of $y_{k,t}$ in \eqref{received sensing data}.

The FIM $\qF_k^{\rm lo}$ can be obtained based on $\qF_k^{\rm ch}$, i.e.,
\begin{equation}
\qF^{\rm lo}_k=\bm{\Gamma}_k^{\qT}\qF_k^{\rm ch}\bm{\Gamma}_k,
\end{equation}
where ${\bm{\Gamma}_k}\in{{\mathbb{R}}^{4\times4}}$  is computed as
\begin{equation}\label{Gamma_k}
[\bm{\Gamma}_k]_{j,m}=\frac{\partial [\bm{\xi}^{\rm ch}_k]_j}{\partial [\bm{\xi}^{\rm lo}_k]_m}.
\end{equation}
Note that the calculation of matrices ${\bf F}_k^{\rm ch}$ and ${\bf \Gamma}_k$ is detailed in Appendix A. Furthermore, the position error matrix is calculated as
\begin{equation}
{\bf \Psi}_k=\left[(\qF^{\rm lo}_k)^{-1}\right]_{{1:2,1:2}}.
\end{equation}
Then, the CRB for ${\bf l}_{k}$ is represented as
\begin{equation}
{\rm CRB}_{{\bf l}_{k}} = \sqrt{{\rm {tr}}{\left( {\bf \Psi}_{k} \right)}}.
\end{equation}
Similarly, the CRB values for ${{\varphi}_{\it k}}$ and ${{\theta}_{\it k}}$ in the channel parameters can be calculated using $\left[(\qF^{\rm ch}_k)^{-1}\right]_{1,1}$ and $\left[(\qF^{\rm ch}_k)^{-1}\right]_{2,2}$, respectively.

In the following, we will introduce the 2D-IFFT based location estimation method \cite{keykhosravi2021siso}.

\subsection{Location Estimation }





We localize the $k$-th UE  by estimating only the AoD from the RIS to the $k$-th UE. To estimate ${\bm \psi}_{\it k}$, first, we define
\begin{equation}
\bar{\bm y}_k=[\bar{y}_{k,1},\bar{y}_{k,2},\ldots,\bar{y}_{k,\tau_{\rm P}B}]^\qT,
\end{equation}
where $\bar{y}_{k,t}={\varrho_k}{[{{\bm \phi}({\bm \psi}_{k})}]}_t$,  ${\varrho_k}={\beta_0}{\beta_k}{({{\bm a}({\bm \psi}_{\rm B})})}^{\qT}{\bf s}$ is an unknown complex scalar in the duration of $\tau_{\rm P}$, and ${[{{\bm \phi}({\bm \psi}_{k})}]}_t$  is  the  $t$-th element of  ${{\bm \phi}({\bm \psi}_{k})}$,  given as
\begin{equation}
[{\bm \phi}({\bm \psi}_{k})]_t={({\bm a}_{\rm x}({\bm \psi}_k))}^{\qT}({{\bm \Lambda_t}}{\odot}{{\bf A}({\bm \psi}_{\rm R})}){\bm a}_{\rm z}({\bm \psi}_k),
\end{equation}
in which  ${\bf A}({\bm \psi}_{\rm R})={{\bm a}_{\rm x}}({\bm \psi}_{\rm R}){({{{{\bm a}_{\rm z}}({\bm \psi}_{\rm R})}})}^{\qT}$, ${\bm \Lambda_t}$ is ${\bm \Lambda}$ at time $t$. Next, we perform the 2D-IFFT operation as follows
\begin{equation}\label{ifft}
\hat{\bm \Lambda}_t={\bf \Gamma}_{\rm x}({{\bm \Lambda}}{\odot}{{\bf A}({\bm \psi}_{\rm R})}){\bf \Gamma}^{\qT}_{\rm z},
\end{equation}
where  ${\bf \Gamma}_{\rm x}\in{\mathbb{C}}^{{M}_{{\rm F}, {\rm x}}{\times}{M_{\rm R,x}}}$ and ${\bf \Gamma}_{\rm z}\in{\mathbb{C}}^{{M}_{{\rm F},{\rm z}}{\times}{M_{\rm R,z}}}$ are the 2D-IFFT matrices whose elements are given as 
$[{\bf \Gamma}_{\rm x}]_{m,k}={\frac{1}{{M}_{{\rm F},{\rm x}}}}e^{j2{\pi}mk/{M}_{{\rm F},{\rm x}}}$ and
$[{\bf \Gamma}_{\rm z}]_{n,l}={\frac{1}{{M}_{{\rm F},{\rm z}}}}e^{j2{\pi}nl/{M}_{{\rm F},{\rm z}}}$, respectively,  $m\in\{0,1,\ldots,{M}_{{\rm F},{\rm x}}-1\}$, $n\in\{0,1,\ldots,{{M}_{{\rm F},{\rm z}}-1}\}$, $k\in\{0,1,\ldots,{{M}_{{\rm R},{\rm x}}-1}\}$, $l\in\{0,1,\ldots,{{M}_{{\rm R},{\rm z}}-1}\}$, and ${M}_{{\rm F},{\rm x}}$ and ${M}_{{\rm F},{\rm z}}$ represent the transformed lengths of the IFFT in the 2D direction. Note that the operation in \eqref{ifft} can be done offline because $\hat{\bm \Lambda}$ does not rely on the received signals online, where $[\hat{\bm \Lambda}]_{m,n}=\big[[\hat{\bm \Lambda}_1]_{m,n},[\hat{\bm \Lambda}_2]_{m,n},\ldots,[\hat{\bm \Lambda}_{{{\tau}_{\rm P}}{B}}]_{m,n}\big]^{\qT}$. 

To narrow down the search for the parameters to be estimated, we assume that there exist integers $ m$ and $n$, $0\leq m\leq {{M}_{{\rm F},{\rm x}}}-1, 0\leq n\leq {{M}_{{\rm F},{\rm z}}}-1$, such that 
\begin{subequations}\label{kthetak}
\begin{align}
[{\bm \mu({\bm \psi}_{\it k})}]_{1}d& {\equiv}2{\pi}m /{{{M}_{{\rm F},{\rm x}}}}  \mod{2\pi},\\
[{\bm \mu({\bm \psi}_{\it k})}]_{3}d& {\equiv}2{\pi}n /{{{M}_{{\rm F},{\rm z}}}} \mod{2\pi}.
\end{align}
\end{subequations}
Furthermore, we define
\begin{equation}
\varepsilon(m,n)={\left\|{{{\bm y}_k}-{f([{\hat{\bf \Lambda}}]_{m,n})}{[{\hat{\bf \Lambda}}]_{m,n}}}\right\|}^2,
\end{equation}
where $[{\hat{\bf \Lambda}}]_{m,n}={\iota_k}{{\bm \phi}({\bm \psi}_{k})}$, $\iota_k$ is a complex scalar (${\varrho_k}\neq \iota_k$ in general), and $f(\bm x)$ reflects the degree of similarity between the two vectors of the same length and is defined as
\begin{equation}
f(\bm x)=\frac{{\bm x}^{\rm H}{{\bm y}_k}}{{\bm x}^{\rm H}{\bm x}}.
\end{equation}
Thus, we need to find $\hat{m}$ and $\hat{n}$ such that 
\begin{equation}
  [\hat{m},\hat{n}]={\mathop{\arg\min}_{m,n}}\ {\varepsilon(m,n)}.  
\end{equation}

Then, the quadratic interpolation technique is employed to obtain better solutions to $\hat{m}$ and $\hat{n}$, i.e.,
\begin{subequations}
\begin{align}
\tilde{m}=\hat{m}+\frac{\varepsilon(\hat{m}-1,n)-\varepsilon(\hat{m}+1,n)}{2[\varepsilon(\hat{m}-1,n)+\varepsilon(\hat{m}+1,n)-2\varepsilon(\hat{m},n)]},\\
\tilde{n}=\hat{n}+\frac{\varepsilon(\hat{n}-1,m)-\varepsilon(m,\hat{n}+1)}{2[\varepsilon(m,\hat{n}-1)+\varepsilon(m,\hat{n}+1)-2\varepsilon(m,\hat{n})]}.
\end{align}
\end{subequations}
Next, we substitute $\tilde{m}$ and $\tilde{n}$ into \eqref{kthetak} to calculate ${[{\bm \omega}(\bm \psi)]}_1$ and ${[{\bm \omega}(\bm \psi)]}_3$. It is obvious that $\varphi_k,\theta_k\in{(0,\pi)}$ in practice because each UE cannot receive signals from the RIS if $\varphi_k\in{(\pi,2\pi)}$.
Then, we calculate $\hat{{\bm \psi}}_k$ using
\begin{equation}
\begin{cases}
{{\hat{\varphi}_k}}&={\rm atan2}\big({[{\bm \omega}(\hat{{\bm \psi}}_k)]}_2,{[{\bm \omega}(\hat{{\bm \psi}}_k)]}_1\big),\\
{{\hat{\theta}}_k}&={\rm acos}\big({[{\bm \omega}(\hat{{\bm \psi}}_k)]}_3\big),
\end{cases}
\end{equation}
where 
\begin{equation}
{[{\bm \omega}(\hat{{\bm \psi}}_k)]}_2=\sqrt{1-{\left({[{\bm \omega}(\hat{{\bm \psi}}_k)]}_1\right)}^2-{\left({[{\bm \omega}(\hat{{\bm \psi}}_k)]}_3\right)}^2}.
\end{equation}
Finally, we improve the estimation accuracy according to 
\begin{equation}\label{theta_best}
{\tilde{\bm \psi}_k}={\mathop{\arg\min}\limits_{{{\bm \psi}_k}}}{\left\|{{{\bm y}_k}-{f({{\bm \phi}({{\bm \psi}}_{k})})}{{{\bm \phi}({{\bm \psi}}_{k})}}}\right\|},
\end{equation}
which can be solved using the quasi-Newton algorithm with the initial iteration value $\hat{{\bm \psi}}_k$. 

With the estimated AoD  ${\tilde{\bm \psi}_k}$, we can calculate the $k$-th UE's position based on the geometric relationship or the prior probability of the $k$-th UE's location ${\hat{{\bf l}}}_{\it k}$. Specifically, we obtain ${\bm \omega}({\tilde{\bm \psi}_k})$ according to \eqref{omega_vec}  and calculate the distance between the $k$-th UE and the RIS as
\begin{equation}
d({\tilde{\bm \psi}_k})=\left\|\hat{\bf l}_{k}-{\bf l}_{\rm R}\right\|=\left| \frac{[{\bf l}_{k}]_3-[{\bf l}_{\rm R}]_3}{\cos({\tilde{\theta}_{\it k}})} \right|,
\end{equation}
where  $[{\bf l}_{k}]_3$ is the height  of UE $k$ which is fixed and known a priori.
Finally,  ${\bf l}_{k}$ is estimated as
\begin{equation}
\hat{\bf l}_k = {\bf l}_{\rm R} + d({\tilde{\bm \psi}_k}){\bm \omega}({\tilde{\bm \psi}_k}).
\end{equation}

\section{Superimposed Pilot and Data Transmission} \label{section: SP}
In this section, we will introduce the uplink transmission with a superimposed pilot and data scheme. Specifically, the channel estimation is first performed with the location sensing result ${\hat{{\bf l}}}_{k}$, then we derive the closed form of the achievable rate in the uplink transmission using channel estimation result. Finally, to maximize the weighted achievable sum rate, we adopt the genetic algorithm to optimize the RIS phase profile.  

\subsection{Ergodic Achievable Rate}
In the same location coherence interval, the UE locations are assumed to be fixed and we analyze the transmission in a channel coherence interval. For the convenience of expression, we remove the interval index in the following discussion without causing ambiguity. 

Denote by $\bm{\varphi}_k\in\mathbb{C}^{\tau_{\rm C}\times 1}$  the  pilot vector for UE $k$ with $\bm{\varphi}^{\rm H}_k\bm{\varphi}_k=\tau_{\rm C},\tau_{\rm C}\geq{\it K}$ and $\bm{\varphi}^{\rm H}_k\bm{\varphi}_{k^{\prime}}=0, k\neq k^{\prime}$. Then, the received signal at the BS during one transmission period is given as
\begin{equation}
\mathbf{Y} = \sum\limits_{k = 1}^K {{\qg_k}\left( {\sqrt {{q_k}} \bm{\varphi} _k^{\rm H} + \sqrt {{p_k}} \bm{\nu}_k^{\rm H}} \right)}  + \bm{\Gamma},
\end{equation}
where $ \qg_k$ is the uplink channel from the $k$-th UE to the BS with $ \qg_k=\qh_{{\rm B},{\it k}}$ due to the uplink-downlink channel reciprocity, $q_k$ and $p_k$ are the normalized signal-to-noise ratios (SNRs) of pilot and data symbols of the $k$-th UE, respectively, ${\bm\nu}_k$ is the data symbols of the $k$-th UE that follows the distribution  ${\mathcal{CN}}(0,{\bf I}_{{\tau}_{\rm C}})$, and $\bm{\Gamma}\in{\mathbb{C}}^{{M_{\rm B}}\times{\tau_{\rm C}}}$ denotes the AWGN matrix whose element follows ${\mathcal{CN}}(0,1)$.

Next, to estimate the uplink channel information, the received signal at the BS is de-spread by multiplying $\bf Y$ with ${\bm {\varphi}}_k/{\sqrt{\tau_{\rm C}}}$, leading to
\begin{equation}
\begin{split}
{\mathbf{y}_k} & = \mathbf{Y}\frac{{{\bm{\varphi} _k}}}{{\sqrt {{\tau_{\rm C}}} }} \\
& = \sqrt {{q_k}{\tau_{\rm C}}} {\mathbf{g}_k} + \sum\limits_{i = 1}^K {\sqrt {\frac{{{p_i}}}{{{\tau _{\rm C}}}}} {\mathbf{g}_i}\bm{\nu}_i^{\rm H}{\bm{\varphi} _k}}  + \bm{\Gamma}\frac{{{\bm{\varphi} _k}}}{{\sqrt {{\tau _{\rm C}}} }}.
\end{split}
\end{equation}
Considering that ${\bf y}_k$ does not follow a Gaussian distribution due to the interference from data symbols. Thus,  the linear minimum mean squared error (LMMSE) metric is adopted to estimate the  corresponding channel as $\hat{\mathbf{g}}_{k}=c_{k} \mathbf{y}_{k}$,  where
\begin{equation}
c_{k}=\frac{\sqrt{q_{k} \tau_{\rm C}} \chi_{k}}{q_{k} \tau_{\rm C} \chi_{k}+\sum\limits_{i=1}^{K} p_{i} \chi_{i}+1} ,
\end{equation}
\begin{equation}\label{xik}
  {\chi _k} = {\rho_{\rm B,\it k}}( {{\varepsilon _0} {\varepsilon _k}{{| {{f_k}} |}^2} + ( {{\varepsilon _0}  + {\varepsilon _k} + 1} ){M_{\rm R}}} ) , 
\end{equation}
and
\begin{equation}
  f_k=   {{\bm a}^{\rm H}({\bm \psi}_{\rm R}}){\rm {diag}}({\rm {vec}}({\bf \Lambda})){{\bm a}}({\bm \psi}_k)  . 
 \end{equation}

In the following, we adopt a maximum ratio combining (MRC) receiver to detect the data from all UEs, i.e.,
\begin{equation}
  \begin{aligned}
  {\hat{\bm \nu}}^{\rm H}_k & = {{\hat{\bf g}}^{\rm H}_k} \left({\bf Y}-\sum_{i=1}^{K} {\sqrt{q_{i}} \hat{\mathbf{g}}_{i} {\bm \varphi}_{i}^{\rm H}}\right)  \cr
  &  = {\hat{\qg}^{\rm H}_k}\sum\limits_{i = 1}^K \left( {\sqrt {{q_i}} {\bm \epsilon}_i \bm{\varphi} _i^{\rm H} + \sqrt {{p_i}} {\bf g}_i \bm{\nu}_i^{\rm H} + \bm{\Gamma}} \right),
  \end{aligned}
\end{equation}
where ${\bm \epsilon}_{k} = {\bf g}_k-{\hat{\bf g}}_k$.
Then, using the use-and-then-forget technique to the received signal, we obtain
\begin{equation}
\hat{ \bm {\nu}}_{k}^{\rm H}= \qz^{\rm H}_{1,k} +\qz^{\rm H}_{2,k}+ \qz^{\rm H}_{3,k},
\end{equation}
where
\begin{subequations}
  \begin{align}
  &\qz^{\rm H}_{1,k} = \sqrt{p_{k}} \lambda_{k} \chi_{k}^{-1} \mathbb{E}\left\{\mathbf{g}_{k}^{\rm H} \mathbf{g}_{k}\right\} \bm{\nu}^{\rm H}_{k}, \\
  &\qz^{\rm H}_{2,k}  = \sqrt{p_{k}} \lambda_{k} \chi_{k}^{-1}\left(\mathbf{g}_{k}^{\rm H} \mathbf{g}_{k}-\mathbb{E}\left\{\mathbf{g}_{k}^{\rm H} \mathbf{g}_{k}\right\}\right) \bm{\nu}^{\rm H}_{k} ,\\
  &\qz^{\rm H}_{3,k} = \qz_{3,k}^{\rm (1)} + \qz_{3,k}^{\rm (2)}+ \qz_{3,k}^{\rm (3)}+ \qz_{3,k}^{\rm (4)},
  \end{align}
  \end{subequations}
with $\lambda_{k}=\sqrt{q_{k} \tau_{\rm C}} \chi_{k} c_{k}$, $\qz_{3,k}^{\rm (1)} = \sqrt{p_{k}} \bar {\mathbf{g}}_{k}^{\rm H} \qg_{k} \bm{\nu}^{\rm H}_k,\qz_{3,k}^{\rm (2)}  = \sqrt{p_{i}} \sum_{i \neq k}^{K} \hat{\qg}_{k}^{\rm H} \qg_{i} \bm{\nu}^{\rm H}_{i},\qz_{3,k}^{\rm (3)} = \sum_{i=1}^{K} \sqrt{q_{i}} \hat{\qg}_{k}^{\rm H} \bm{\epsilon}_{i} \bm{\varphi}^{\rm H}_{i},\qz_{3,k}^{\rm (4)}   =\hat{\mathbf{g}}_{k}^{\rm H} \bm{\Gamma}$,
and  $\bar{\qg}_{k}=\hat{\qg}_{k}-c_k\sqrt{q_{k} \tau_{\rm C}}\qg_{k}$.
Then the  effective signal-to-interference-plus-noise ratio (SINR) of the $k$-th UE is
\begin{equation}\label{SINR}
\gamma_{k}=\frac{|\qz_{1,k}|^2}
{
\mathbb{E}|\qz_{2,k}|^2 
+\mathbb{E}\left\{\|\qz_{3,k}-\mathbb{E}\{|\qz_{3,k}|\}\|^{2}\right\}.}
\end{equation}
\textit{\begin{theorem}\label{theorem1}
Considering the uplink SP transmission where the  MRC receiver is adopted.  The lower bound of the ergodic achievable rate of  the $k$-th UE  can be given as
$R_k=\log_2{(1+\bar{\gamma}_k)}$, where
\begin{equation}\label{deterministic sinr}
\bar{\gamma}_{k}=\frac{{M^{2}_{\rm B}} \lambda_{k}^{2} p_{k} \tau_{\rm C}}{\sum^4_{i=0}\Pi_i},
\end{equation}
in which $\Pi_0={p_k\tau_C M_{\rm B} \lambda_{k}^{2}(\chi_{k}^{-2} \Delta_{k}-M_{\rm B})}$,
$\Pi_1=\sum_{i=1}^4\Pi_{1i}$,
\begin{equation}
\begin{split}
\Pi_{11}&=M_{\rm B} c_{k}^{2} p_{k}^{2} (\tau_{\rm C}+1) \Delta_{k} +M_{\rm B} c_{k}^{2} \tau_{\rm C} \sum_{i \neq k}^{K} p_{i} p_{k} \Omega_{k i} \\
& +M_{\rm B} c_{k}^{2} \tau_{\rm C} p_{k} \chi_{k},\\
\Pi_{12} & =M_{\rm B} q_{k} c_{k}^{2} \tau_{\rm C}^{2} \sum_{i \neq k}^{K} p_{i} \Omega_{k i} +M_{\rm B} \tau_{\rm C} c_{k}^{2} \sum_{i \neq k}^{K} p_{i}^{2} \Delta_{i} \\
& +M_{\rm B} c_{k}^{2} \sum_{i \neq k}^{K} p_{i}^{2} \Delta_{i}  +M_{\rm B} c_{k}^{2} \sum_{\substack{i \neq k}}^{K} \sum_{\substack{j \neq k \\
j \neq i}}^{K} p_{i} p_{j} \Xi_{i j} \\
& +M_{\rm B} \tau_{\rm C} c_{k}^{2} \sum_{i \neq k}^{K} \sum_{j \neq i}^{K} p_{i} p_{j} \Omega_{i j}  +M_{\rm B} \tau_{\rm C} c_{k}^{2} \sum_{i \neq k}^{K} p_{i} \chi_{i},\\
\Pi_{13}&=M_{\rm B} \tau_{\rm C} \sum_{i=1}^{K} q_{i} \lambda_{k}(\chi_{i}-\lambda_{i}),\\
\Pi_{14}&=M_{\rm B} \lambda_k \tau_{\rm C}+c_k^2 M^2_{\rm B},\\
\Pi_2&=-\big[M^2_{\rm B} p_k^2 c_k^2 \chi_k^2+M^2_{\rm B} c_k^2(\sum_{i \neq k}^K p_i \chi_i)^2+M^2_{\rm B}c^2_k\big],\\
\Pi_3&= -2M^2_{\rm B} c_k^2 \Big(p_k \chi_k \sum_{i \neq k}^K p_i \chi_i  + p_k \chi_k+  \sum_{i \neq k}^K p_i \chi_i\Big),
\end{split}
\end{equation}
and
\begin{equation}
\Pi_4=2M_{\rm B} c_k^2 \Big(p_k \sum_{i \neq k}^K p_i \Xi_{i k}+ M_{\rm B} p_k \chi_k  + M_{\rm B} \sum_{i \neq k}^K p_i \chi_i\Big).
\end{equation}
\end{theorem}}
\begin{IEEEproof}
   The detailed proof is given in Appendix B.
\end{IEEEproof}
\subsection{RIS Phase Profile Optimization}
With the closed-form expression of the achievable rate in \textbf{Theorem 1}, we aim to maximize the weighted sum rate by  optimizing the phase profile of the RIS, i.e.,
\begin{equation}\label{eq:my_equation}
 \begin{split}
 \max_{\bm{\Lambda}}\ &\sum_{k\in\mathcal{K}}\kappa_kR_k \\
 &|[{\bf\Lambda}]_{m_{\rm x},m_{\rm z}}|=1,\forall m_{\rm x},m_{\rm z},
 \end{split}
\end{equation}
where $\kappa_k$ is the weight of UE $k$. Note that the deterministic expression $\bar{\gamma}_k$ in \eqref{deterministic sinr} is very complicated and traditional optimization algorithms are hard to apply. Hence, we resort to GA to solve this problem.  The main idea of GA is to find solutions by simulating the mechanism of evolution in nature \cite{Melanie1998An}. With great parallelism and global searching ability, GA is widely adopted in existing works \cite{Zhi2022Power, Peng2021Analysis, Dai2022Statistical}, where the main steps of GA include population initialization, fitness evaluation, selection, crossover and mutation and the detailed process of GA is omitted due to the limited space.

\section{Simulation Results}\label{section: simulations}
We consider a scenario where a BS is equipped with a UPA  and serves $K=4$ single-antenna UEs,  which are randomly distributed on the $xOy$ plane.  The heights of the UEs, RIS, and BS are $[{\bf l}_{\it k}]_3=0, \forall k$,  $[{\bf l}_{\rm B}]_3=9$m, and $[{\bf l}_{\rm R}]_3=10$ m, respectively. The BS and RIS coordinate systems have the same 3D standard orthogonal basis as the global coordinate system and the RIS is placed at the coordinate origin. Without losing generality, it is assumed $p_k+q_k=P_{\rm U}, \forall k$.

We also assume that the carrier frequency $f_{\rm C}=28$ GHz and the maximum velocity of these UEs is 1 m/s, then we have the corresponding maximum Doppler shift   $f_{\rm D}\approx 93$ Hz and can set the channel coherence interval to be $\tau_{\rm C}=1$ ms \cite{jiang2022optimization}. Given the bandwidth $B$= 100 kHz and $\tau_{\rm L}=1$ s, each location coherent interval has 1000 channel coherent intervals each with 100 symbols. More parameters are listed in Table \ref{table:1} and default values will be used, if not specified.
\begin{table}[h!]
\caption{Parameters used in the simulations.}
\label{table:1}
\centering
\begin{tabular}{ l l l }
  \hline
  Parameter & Symbol & Value \\ [0.5ex]
  \hline
  Sensing interval & ${\tau}_{\rm P}$ & 1 ms \\
  BS location & ${\bf l}_{\rm B}$ & $[5,5,9]^\qT$ \\
  RIS location & ${\bf l}_{\rm R}$ & $[0,0,10]^\qT$ \\
  BS/RIS rotation matrix & ${\bf V}_{\rm B},{\bf V}_{\rm R}$ & $\qI_3$ \\
  BS UPA size & $M_{\rm B,x}\times M_{\rm B,z}$ & $10\times 10$ \\
  RIS UPA size & $M_{\rm R,x}\times M_{\rm R,z}$ & $10\times 10$ \\
  Transmit power of the BS & $P_{\rm B}$ & 500 mW \\
  Transmit power of the UEs & $P_{\rm U}$ & 200 mW \\
  Noise power density & $N_0$ & -174 dBm/Hz \\
  UE’s Noise figure & $n_f$ & 8 dB \\
  Path-loss exponent  & $b$ & 2 \\
  Rician factors & ${\varepsilon}_0={\varepsilon}_k$ & 50 \\
  2-D IFFT dimensions & ${{\bf \Gamma}_{\rm x}}={{\bf \Gamma}_{\rm y}}$ & 256 \\ [1ex]
  \hline
\end{tabular}
\end{table}
\subsection{Location Sensing Performance}
 In this section, we first verify the efficiency of the  2D IFFT estimation algorithm by comparing the root mean square error (RMSE) of the location estimation results with the CRB. We assume all the phase shifts of the RIS elements are drawn from a uniform distribution over $[0,2\pi)$ independently.    To evaluate the estimation performance of the IFFT method, without losing generality, we choose one UE as an example and average on 1000 noisy realizations.

\begin{figure}
  \centering
  \subfigure[]{
    \label{position estimation} 
    \includegraphics[width=0.47\textwidth]{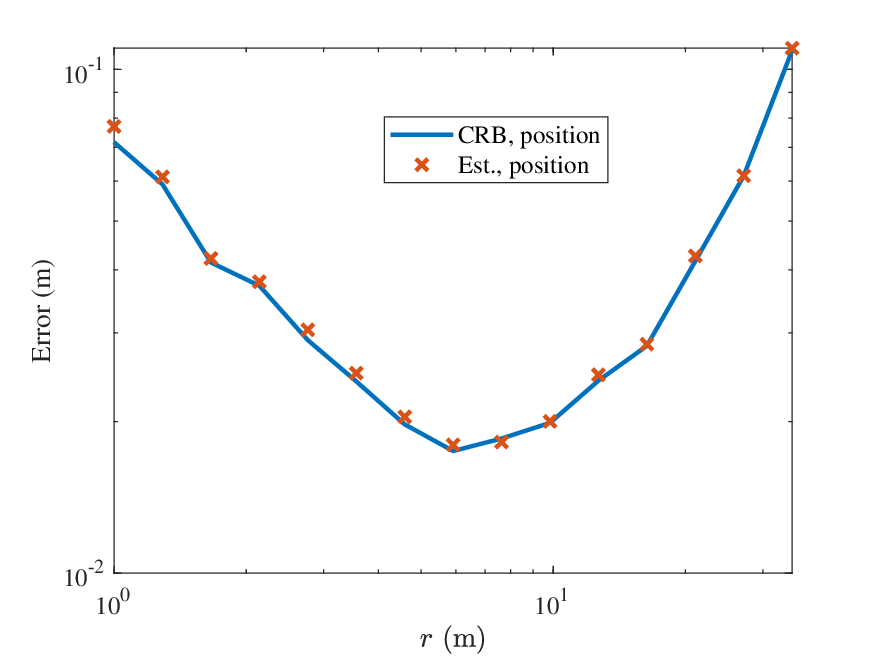}}
  \subfigure[]{
    \label{angle estimation} 
    \includegraphics[width=0.47\textwidth]{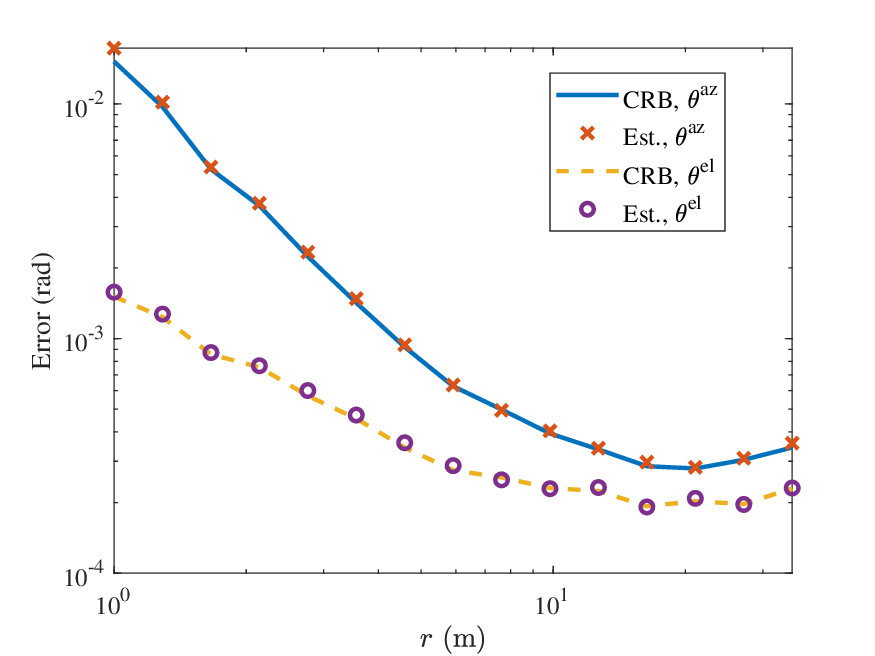}}
  \caption{The estimation error and CRB of (a) UE position and (b) angles. The RIS has a UPA of 64 $\times 64$.}
  \label{position_angle_estimation}
\end{figure}

Figs. \ref{position estimation} and \ref{angle estimation} show the CRB and the estimation errors of the UE position and channel parameters along the direction $[-r/\sqrt{2},r/\sqrt{2},0]$, respectively. It is observed that the estimation error of the adopted IFFT method is very close to the theoretical CRB. We can also find that a small value of $r$ leads to low accuracy of angle estimation. This is because the UE is too close to the RIS and can be considered to be almost beneath the RIS especially when $r$ is very small. In such a case, the position error is relatively large. The positioning accuracy at the level of submeter can be achieved when $r<30$ m.  In addition, we can observe that the estimation performance degrades with $r>15$ m  due to the decrease of the SNR.

\begin{figure}
\centering
\includegraphics[width=0.47\textwidth]{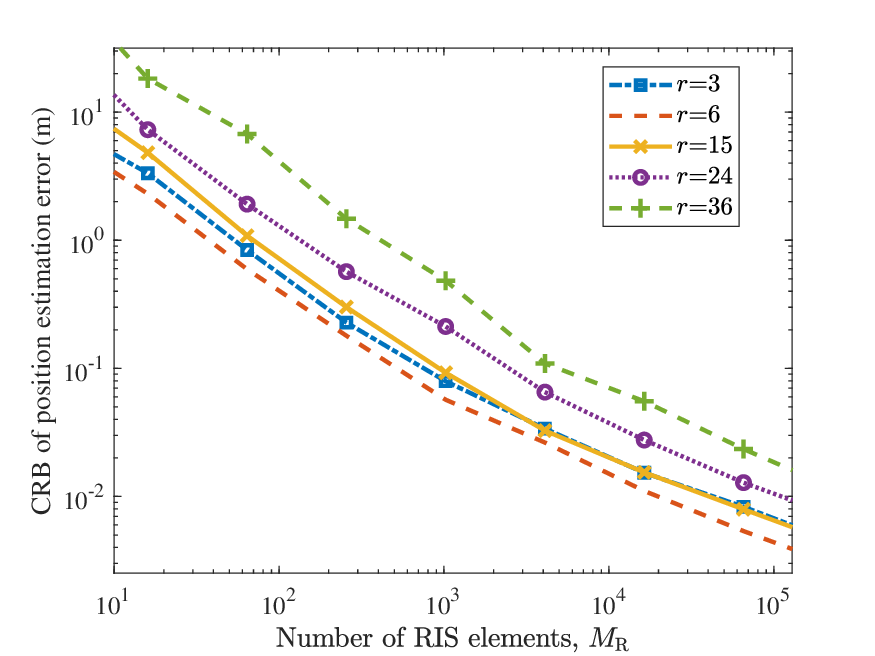}
\caption{The CRB of position estimation error with respect to the number of RIS elements.}
\label{PEB}
\end{figure}

Fig. \ref{PEB} illustrates the CRB of position estimation error over the number of RIS elements ($M_{\rm R}$) with the assumption of $M_{{\rm R},x}=M_{{\rm R},z}$. As observed, more RIS elements lead to a smaller CRB of position estimation error. We also find that the position estimation error increases with the increase of the sensed area size due to the decrease of SNR. However,  we can improve the location estimation accuracy by increasing the number of RIS elements. 

\begin{figure}
  \centering
  \subfigure[]{
    \label{postionerror} 
    \includegraphics[width=0.47\textwidth]{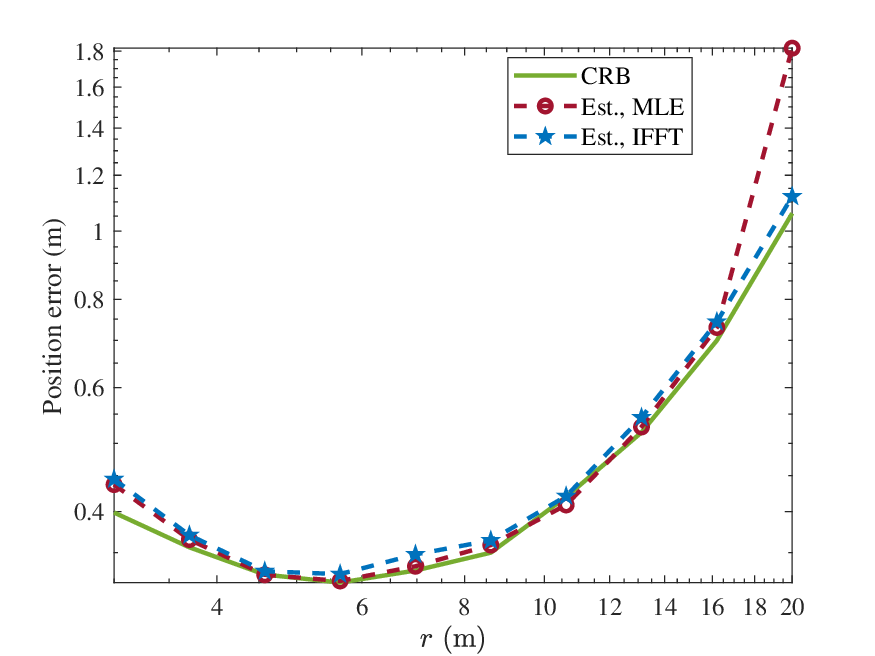}}
  \subfigure[]{
    \label{running time} 
    \includegraphics[width=0.47\textwidth]{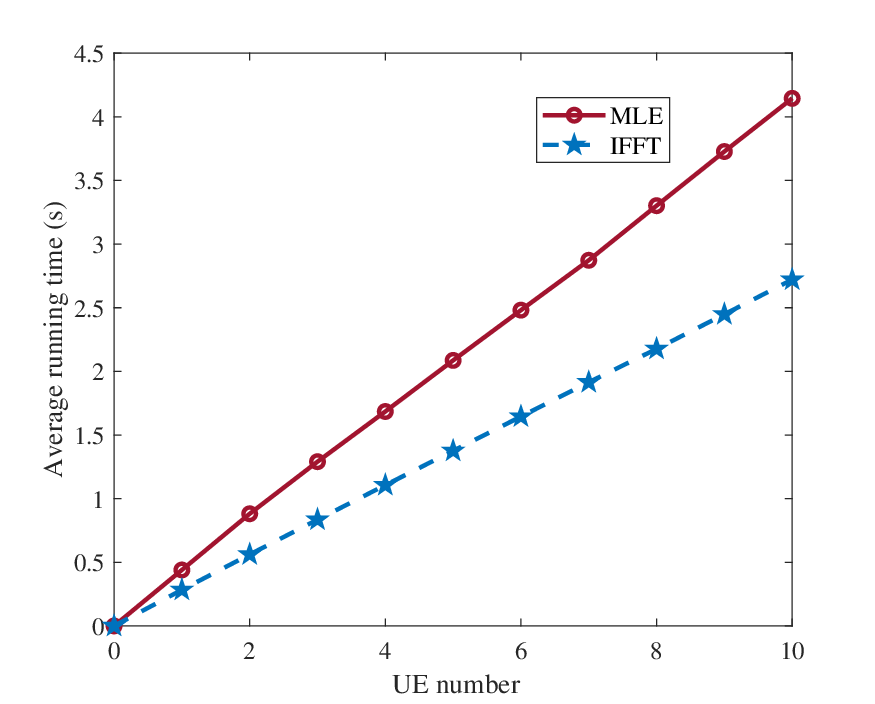}}
  \caption{Comparison of the IFFT and MLE algorithms: (a) position error and (b) average running time per sample. }
  \label{comp_ifft_mle}
\end{figure}

Fig. \ref{comp_ifft_mle} compares the IFFT algorithm with the maximum likelihood estimation (MLE) algorithm \cite{fascista2022ris} in terms of positioning error and algorithm running time. We can find the performance of the MLE algorithm degrades more sharply when the sensing area is large. Besides, the MLE algorithm has high computational complexity. Thus, the IFFT algorithm is chosen as the base of the following simulation results.

\subsection{Uplink SP Transmission Performance}
Here, we assume the UEs are evenly distributed within a 17m$\times$17m rectangular area with $x\in[-3{\rm m},-20{\rm m}]$, $y\in[3{\rm m},20{\rm m}]$  and $\kappa_k=1,\forall k$.  Besides, for convenience of comparison, we also assume $p_k=p, q_k=q, \forall k$ and define  $\eta=p/(p+q)$ as the power allocation factor. To calculate the analytical results, we set the GA population size, the elite number, the number of crossover parents, the number of mutation parents, and the mutation probability be 200, 10, 160, 20, 0.1, respectively.

\begin{figure}
\centering
\includegraphics[width=0.47\textwidth]{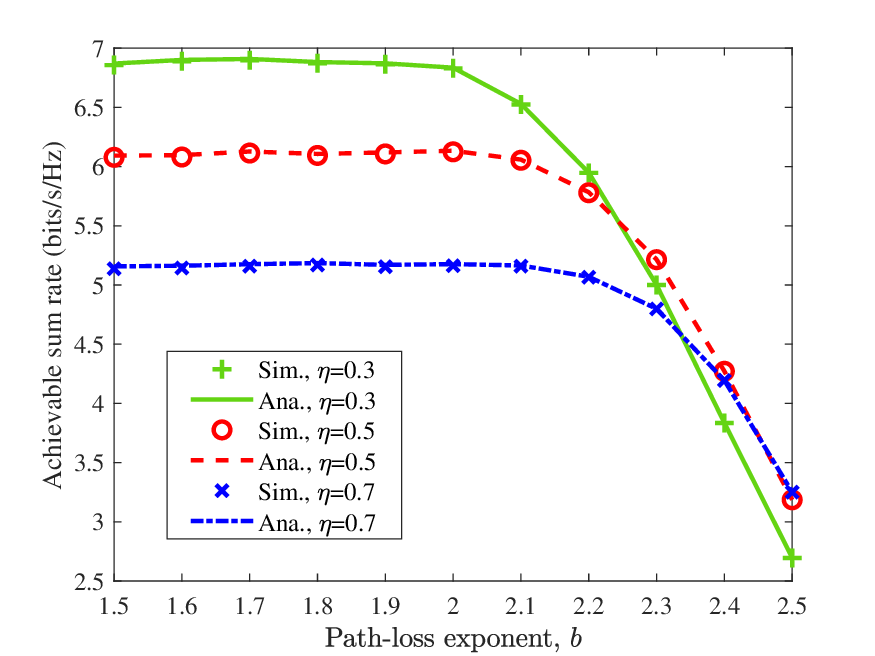}
\caption{Comparison of simulation and analytical results with $K=4$.}
\label{sum_rate_path_loss}
\end{figure}

\begin{figure}
\centering
\includegraphics[width=0.47\textwidth]{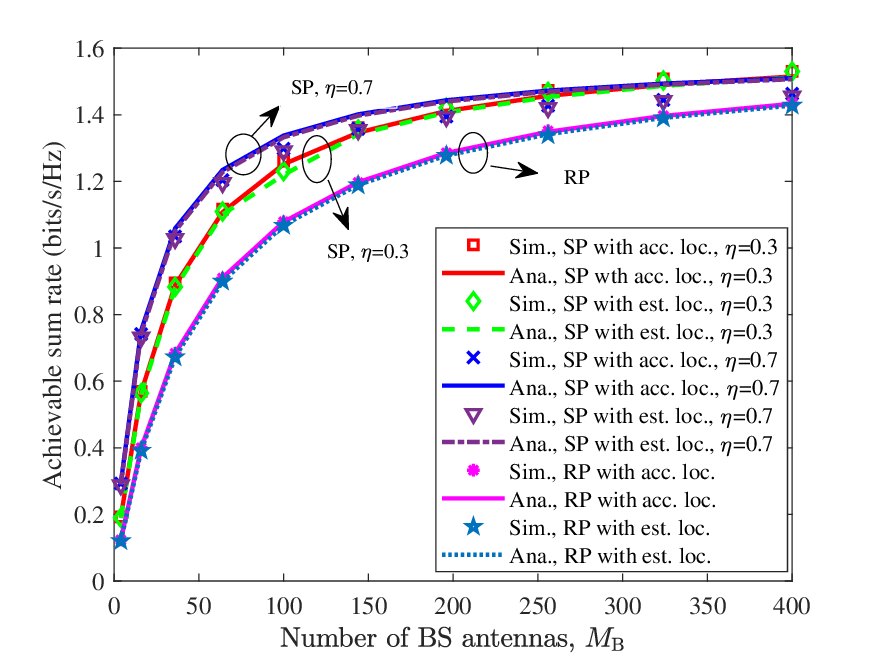}
\caption{Comparison of simulation and analytical results.  The RIS  has a UPA of $2\times 2$ and $K=4$ UEs are considered.}
\label{sim_ana_comp}
\end{figure}

\begin{figure}
\centering
\includegraphics[width=0.47\textwidth]{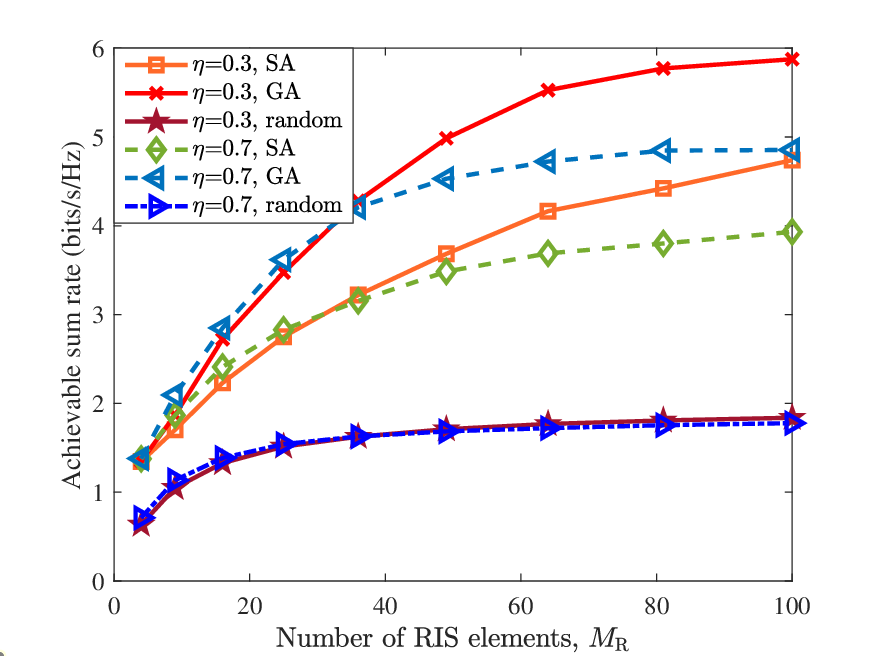}
\caption{Achievable sum rate with respect to the number of RIS elements. Here, we assume $M_{{\rm R},x}=M_{{\rm R},z}$ and  $M_{{\rm B},x}=M_{{\rm B},z}$. }
\label{SR_RIS}
\end{figure}

\begin{figure}
\centering
\includegraphics[width=0.47\textwidth]{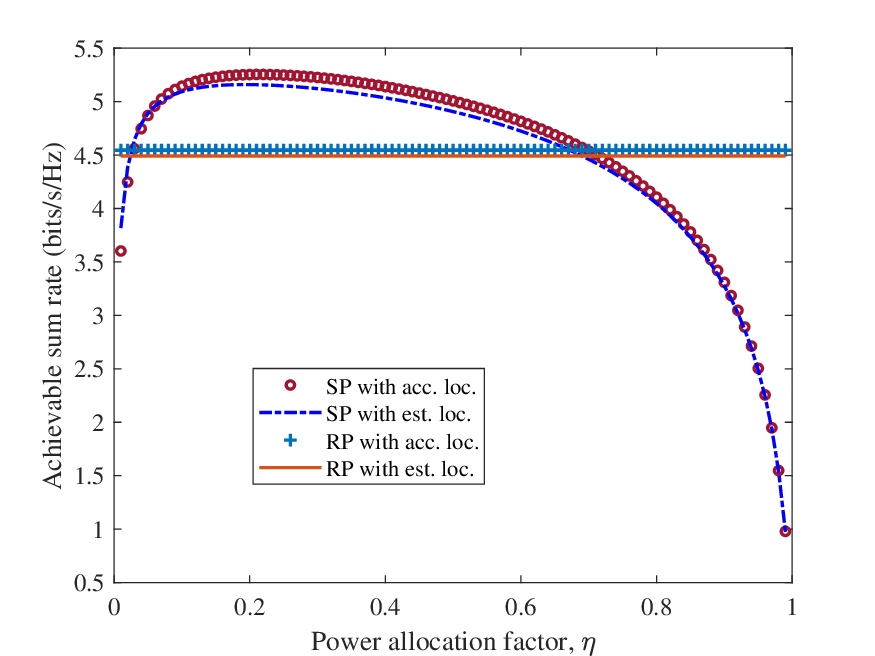}
\caption{Effect of power allocation on achievable sum rate. The RIS has a UPA of  $20\times20$ and $K=16$ UEs are considered.}
\label{SR_power}
\end{figure}

Fig. \ref{sum_rate_path_loss} shows the simulation (marked as ``Sim.") results based on the Monte-Carlo method and analytical (marked as ``Ana.")  results based on \textbf{Theorem 1}. As observed, the curves of simulation and analytical results are tight, which validates the accuracy of the lower bound of the  achievable rate in \textbf{Theorem 1}.  
Furthermore, Fig. \ref{sim_ana_comp} compares  the SP and RP schemes considering  accurate locations (marked as ``acc. loc.")  and estimated locations (marked as ``est. loc.").   We  find that the obtained results  using the estimated locations are close to that using the accurate locations in different cases because of high estimation accuracy of UE locations. We also observe that different power allocation factors lead to different achievable rate performance, which suggests that we can improve the system performance with an optimized power allocation between the data and pilots.

Fig. \ref{SR_RIS} shows the achievable sum rate with respect to the number of the RIS elements ($M_{\rm R}$) with the assumption of $M_{{\rm R},x}=M_{{\rm R},z}$ and  $M_{{\rm B},x}=M_{{\rm B},z}$.  Here, we introduce two baselines, i.e., random RIS phase and optimized RIS phase using the simulation annealing (SA) algorithm, for comparison. Specifically, the random RIS phase is obtained as the average of 10000 samples. The main idea of the SA algorithm is to simulate the cooling process of materials, performing a random search in the solution space to find the global optimum. The SA algorithm can be seen as the special case of the GA algorithm with one population. 
We set  the initial temperature and the cooling rate of the SA algorithm be 1000 and 0.99, respectively.   As observed, More RIS elements lead to a higher achievable sum rate. Moreover, the sum rate of the optimized RIS phase is much larger than that of the random RIS phase, which reflects the necessity of RIS phase optimization. According to the results of  Figs. \ref{PEB} and \ref{SR_RIS}, we can conclude that the  increase of the RIS elements can improve the position sensing accuracy, but also  the achievable sum rate. 
  
Finally, Fig. \ref{SR_power} demonstrates the effects of different power allocation schemes. We can find that there is a peak that can maximize the achievable sum rate, which suggests we can jointly optimize RIS phase profile (e.g., using GA) and power allocation among data and pilots (e.g., using the bisection method).   

\section{Conclusions}\label{section: conclusions}
In this work, we first designed a frame structure of RIS-assisted SP transmission protocol composed of several location coherence intervals each with pure-pilot transmission duration and data-pilot transmission duration. The pure-pilot transmission duration is used to estimate UE locations using the IFFT algorithm.  In contrast, the data-pilot transmission duration is time-slotted match the channel coherence time, and the data and pilot signals are transmitted simultaneously. 
We conducted the FIM analysis and derived the CRB of position estimation error.  Furthermore, we derived the closed-form lower bound of the ergodic achievable rate, based on which we optimized the phase profile of the RIS to maximize the achievable sum rate using the genetic algorithm. Finally, numerical results validated the accuracy of the  UE position estimation using the IFFT algorithm and the superiority of the proposed SP transmission scheme as compared to the RP scheme.

\begin{appendices}
\section{Calculation of ${\bf F}_k^{\rm ch}$ and ${\bm \Gamma}_k$}
We first calculate ${\bf F}_k^{\rm ch}$. In \eqref{FIM}, we have
\begin{subequations}
\begin{align}
{\frac{\partial {{\bar{y}}_{k,t}}}{\partial {\varphi_k}}}&={\bigg(}{\frac{\partial {{\bar{\bf h}}_{k}}}{\partial {\varphi_k}}}{\bigg)}{\rm diag}({\rm {vec}}({\bf \Lambda})){\bar{\bf H}_0}\qs,\\
{\frac{\partial {{\bar{y}}_{k,t}}}{\partial {\theta_k}}}&={\bigg(}{\frac{\partial {{\bar{\bf h}}_{k}}}{\partial {\theta_k}}}{\bigg)}{\rm diag}({\rm {vec}}({\bf \Lambda})){\bar{\qH}_0}\qs,
\end{align}
\end{subequations}
where
\begin{subequations}
\begin{align}
{\frac{\partial {{\bar{\bf h}}_{k}}}{\partial {\varphi_k}}}&={\beta_k}{\bigg(}{\frac{\partial {{\bm a}{({\bm \psi}_k})}}{\partial {\varphi_k}}}{\bigg)}^{\qT},\\
{\frac{\partial {{\bar{\bf h}}_{k}}}{\partial {\theta_k}}}&={\beta_k}{\bigg(}{\frac{\partial {{\bm a}{({\bm \psi}_k})}}{\partial {\theta_k}}}{\bigg)}^{\qT},\\
{\frac{\partial {{{\bm a}{({\bm \psi}_k})}}}{\partial {\varphi_k}}}&={{\bm a}{({\bm \psi}_k})}{\odot}{\bigg(}-j{\Big(}{\frac{\partial {{\bm \mu}{({\bm \psi}_k})}}{\partial {\varphi_k}}}{\Big)}^{\qT}{\bf N}{\bigg)},\\
{\frac{\partial {{{\bm a}{({\bm \psi}_k})}}}{\partial {\theta_k}}}&={{\bm a}{({\bm \psi}_k})}{\odot}{\bigg(}-j{\Big(}{\frac{\partial {{\bm \mu}{({\bm \psi}_k})}}{\partial {\theta_k}}}{\Big)}^{\qT}{\bf N}{\bigg)},\\
{\frac{\partial {{\bm \mu}({\bm \psi}_k)}}{\partial \varphi_k}}&={-\frac{2\pi}{\lambda_{\rm c}}}{\frac{\partial {{\bm \omega}({\bm \psi}_k)}}{\partial \varphi_k}}\\
{\frac{\partial {{\bm \mu}({\bm \psi}_k)}}{\partial \theta_k}}&={-\frac{2\pi}{\lambda_{\rm c}}}{\frac{\partial {{\bm \omega}({\bm \psi}_k)}}{\partial \theta_k}},\\
{\frac{\partial {{\bm \omega}{({\bm \psi}_k})}}{\partial {\varphi_k}}}&={[{{-{\sin{\theta_k}}{\sin{\varphi_k}}},{-{\sin{\theta_k }}{\cos{\varphi_k}}},0}]}^{\qT},\\
{\frac{\partial {{\bm \omega}{({\bm \psi}_k})}}{\partial {\theta_k}}}&={[{{{\cos{\theta_k}}{\cos{\varphi_k}}},{{\cos{\theta_k}}{\sin{\varphi_k}}},-{{\sin}{\theta_k}}}]}^{\qT},
\end{align}
\end{subequations}
and
\begin{equation}
[\frac{\partial {{\bar{y}}_{k,t}}}{\partial {\mathfrak{R}(\beta_k)}},\frac{\partial {{\bar{y}}_{k,t}}}{\partial {\mathfrak{J}(\beta_k)}}]={({{\bm a}({\bm \psi}_{\it k})})}^{\qT}{\rm {diag}}{({\rm {vec}}({\bf \Lambda}))}{\bar{\bf H}_0}{\qs}[1,j].
\end{equation}

Then, we calculate ${\bm \Gamma}_k$. In \eqref{Gamma_k}, we have
\begin{subequations}
\begin{align}
&[\frac{\partial {\mathfrak{R}(\beta_k)}}{\partial {\mathfrak{R}(\beta_k)}},\frac{\partial {\mathfrak{J}(\beta_k)}}{\partial {\mathfrak{J}(\beta_k)}},\frac{\partial {\mathfrak{R}(\beta_k)}}{\partial {\mathfrak{J}(\beta_k)}},\frac{\partial {\mathfrak{J}(\beta_k)}}{\partial {\mathfrak{R}(\beta_k)}}]=[1,1,0,0],\\
&\frac{\partial \varphi_k}{\partial \ql_k}=\frac{1}{\Delta p}\left[-[\Delta{\ql_k}]_{2} {[{\bf V}_{\rm R}]}_{1,1:3}+[\Delta{\ql_k}]_{1} {[{\bf V}_{\rm R}]}_{2,1:3}\right],\\
&\frac{\partial \theta_k}{\partial \ql_k}=\frac{-\|\Delta{\ql_k}\|^{2}[{\bf V}_{\rm R}]_{3,1:3}+\left(\ql_k-\ql_{\mathrm{R}}\right)[\Delta{\ql_k}]_{3}}{\|\Delta{\ql_k}\|^{2} \sqrt{\Delta p}},
\end{align}
\end{subequations}
where $\Delta p={\left([\Delta{\ql_k}]_{1}\right)^{2}+\left([\Delta{\ql_k}]_{2}\right)^{2}}$. Besides, the remaining derivatives are zero.

\section{Proof of Theorem 1}



We rewrite $ {\bf g}_k$ as
\begin{equation}
  {\bf g}_k={\bf g}_k^{(1)}+{\bf g}_k^{(2)}+{\bf g}_k^{(3)}+{\bf g}_k^{(4)},
  \end{equation}
where ${\bf g}_k^{(1)}=\bar{\bf H}_0{\bf \Phi}\bar{\bf h}_k$, ${\bf g}_k^{(2)}=\bar{\bf H}_0{\bf \Phi}\tilde{\bf h}_k$, ${\bf g}_k^{(3)}=\tilde{\bf H}_0{\bf \Phi}\bar{\bf h}_k$,
  and $\qg_k^{(4)}=\tilde{\bf H}_0{\bf \Phi}\tilde{\qh}_k$.
Considering that
   \begin{equation}
  \mathbb{E}\{ \|\qg_k\|^2 \} =\sum_{i=1}^4\mathbb{E} \{\| {{\bf g}_k^{(i)}}\|^2 \},
  \end{equation}
where 
  \begin{subequations}
  \begin{align}
  &\mathbb{E} \{\| {{\bf g}_k^{(1)}} \|^2\}={\rho_{\rm B,\it k}} {\varepsilon _0} {\varepsilon _k}{M_{\rm B}}{{\left| {{f_k}} \right|}^2},\\
  &{\mathbb{E}}\{ {\| {\bf g}_k^{(2)} \|}^2 \}={\rho_{\rm B,\it k}} {\varepsilon _0} {M_{\rm B}} {M_{\rm R}},\\
  &{\mathbb{E}}\{ {\| {\bf g}_k^{(3)} \|}^2 \}={\rho_{\rm B,\it k}} {\varepsilon _k} {M_{\rm B}} {M_{\rm R}},\\
  &{\mathbb{E}}\{ {\| {\bf g}_k^{(4)} \|}^2 \}={\rho_{\rm B,\it k}}  {M_{\rm B}} {M_{\rm R}}.
  \end{align}
  \end{subequations}  

Thus we have
 \begin{equation}
  \mathbb{E}\{ \|\qg_k\|^2 \} = {M_{\rm B}}{\chi _k},
  \end{equation}
in  which ${\chi _k}$ is defined as  \eqref{xik}.
We further have $\mathbb{E}\{\left\|\hat{\qg}_{k}\right\|^{2}\}=M_{\rm B} \lambda_{k}$
and $\mathbb{E}\{\left\|{\bm \epsilon}_{k}\right\|^{2}\}=M_{\rm B}  (\chi_{k} -\lambda_{k})$ with ${\bm \epsilon}_{k} = {\bf g}_k-{\hat{\bf g}}_k$.


Based on the above results, we  obtain
\begin{subequations}
\begin{align}
\mathbb{E}\{\left\|\mathbf{g}_k\right\|^4\}&=M_{\rm B} \Delta_k,\\
\mathbb{E}\{\left|\mathbf{g}_k^{\rm H} \mathbf{g}_i\right|^2\}&=M_{\rm B} \Omega_{k i},\\
\mathbb{E}\{\mathbf{g}_k^{\rm H} \mathbf{g}_k \mathbf{g}_i^{\rm H} \mathbf{g}_i\}&=M_{\rm B} \Xi_{k i},
\end{align}
\end{subequations}
where
\begin{equation}
  \begin{aligned}
  \Delta_{k} & ={\rho^{2}_{\rm B,\it k}} \left\{{M_{\rm B}}(\varepsilon_{0} \varepsilon_{k})^{2}|f_{k}|^{4}\right. \\
  & +2 \varepsilon_{0} \varepsilon_{k}|f_{k}|^{2}(2 {M_{\rm B}} {M_{\rm R}} \varepsilon_{0}+{M_{\rm B}} {M_{\rm R}} \varepsilon_{k} \\
  & +{M_{\rm B}} {M_{\rm R}}+2 {M_{\rm B}}+{M_{\rm R}} \varepsilon_{k}+{M_{\rm R}}+2) \\
  & +{M_{\rm B}} {M^{2}_{\rm R}}(2 \varepsilon_{0}^{2}+\varepsilon_{k}^{2}+2 \varepsilon_{0} \varepsilon_{k}+2 \varepsilon_{0}+2 \varepsilon_{k}+1) \\
  & +{M^{2}_{\rm R}}(\varepsilon_{k}^{2}+2 \varepsilon_{0} \varepsilon_{k}+2 \varepsilon_{0}+2 \varepsilon_{k}+1) \\
  & \left.+{M_{\rm B}} {M_{\rm R}}(2 \varepsilon_{0}+2 \varepsilon_{k}+1) +{M_{\rm R}}(2 \varepsilon_{0}+2 \varepsilon_{k}+1)\right\},
  \end{aligned}
  \end{equation}

\begin{equation}
  \begin{aligned}
  \Omega_{k i} & ={\rho_{\rm B,\it i}{\rho_{\rm B,\it k}}} \big\{ M_{\rm B} \varepsilon^{2}_{0} \varepsilon_{k} \varepsilon_{i}|f_{k}|^{2}|f_{i}|^2\\
  & +\varepsilon_{0} \varepsilon_{k}|f_{k}|^{2} \times(\varepsilon_{0} M_{\rm B} M_{\rm R} +M_{\rm R} \varepsilon_{i}+M_{\rm R}+2 M_{\rm B}) \\
  & +\varepsilon_{0} \varepsilon_{i}|f_{i}|^{2} \times(\varepsilon_{0} M_{\rm B} M_{\rm R}+M_{\rm R} \varepsilon_{k}+M_{\rm R}+2 M_{\rm B}) \\
  & +M^{2}_{\rm R} \times\left(M_{\rm B} \varepsilon^{2}_{0}+\varepsilon_{0}(\varepsilon_{i}+\varepsilon_{k}+2)+(\varepsilon_{k}+1)(\varepsilon_{i}+1)\right) \\
  & +M_{\rm B} M_{\rm R}(2 \varepsilon_{0}+\varepsilon_{i}+\varepsilon_{k}+1) +M_{\rm B} \varepsilon_{k} \varepsilon_{i}|{\bar{\bf{h}}}_{k}^{\rm H} {\bar{\bf{h}}}_{i}|^{2} \\
  & +2 M_{\rm B} \varepsilon_{0} \varepsilon_{k} \varepsilon_{i} \mathbf{R e}\{f_{k}^{\rm H} f_{i} {\bar{\qh}}_{i}^{\rm H} {\bar{\qh}}_{k}\}\big\},
  \end{aligned}
\end{equation}
and
\begin{equation}
\begin{split}
   \Xi_{k i}&\!=\! M_{\rm B} \chi_{k} \chi_{i} + {\rho_{\rm B,\it i}}{\rho_{\rm B,\it k}}  (\varepsilon_{k} \varepsilon_{i}|{\bar{\bf{h}}}_{i}^{\rm H} {\bar{\bf{h}}}_{k}|^{2}+M_{\rm R}(\varepsilon_{i}+\varepsilon_{k}+1) \\
   &+ 2{\varepsilon_0}{\varepsilon_k}|f_k |^2 +2{\varepsilon_0}{\varepsilon_i}|f_i|^2+2M_{\rm R} \varepsilon_0).
\end{split}
\end{equation}
We rewrite  the numerator of $\gamma_k$ in \eqref{SINR} as
\begin{equation}\label{numerator}
\begin{split}
|\qz_{1,k}|^2&=p_{k} \lambda^2_k\chi_k^{-2}\tau_{\rm C} \left|\mathbb{E}\{ ||\qg_{k}||^2\}\right|^{2}\\  
&=p_k\tau_{\rm C} M^{2}_{\rm B} \lambda_{k}^{2}.
\end{split}
\end{equation} 
The denominator in (\ref{SINR}) has two parts and will be dealt with separately in the following. We first address the left part as
\begin{equation}\label{2}
\begin{split}
\mathbb{E}|\qz_{2,k}|^2 &=p_k\tau_{\rm C} \operatorname{Var}(\lambda_{k} \chi_{k}^{-1} \qg_{k}^{\rm H} \qg_{k}) \\
&=p_k\tau_{\rm C} \mathbb{E}\{|\lambda_{k} \chi_{k}^{-1} \qg_{k}^{\rm H} \qg_{k}|^{2}\} \\
&-p_k\tau_{\rm C} |\lambda_{k} \chi_{k}^{-1} \mathbb{E}\{\qg_{k}^{\rm H} \qg_{k}\}|^{2}.
\end{split}
\end{equation}
Because of
\begin{equation}
\begin{split}
\mathbb{E}\left\{\left|\lambda_{k} \chi_{k}^{-1} \mathbf{g}_{k}^{\rm H} \mathbf{g}_{k}\right|^{2}\right\}& =\lambda_{k}^{2} \chi_{k}^{-2} \mathbb{E}\left\{\left\|\mathbf{g}_{k}\right\|^{4}\right\} \\
& =\lambda_{k}^{2} \chi_{k}^{-2} M_{\rm B} \Delta_{k}, 
\end{split}
\end{equation}
and 
\begin{equation}
|\lambda_{k} \chi_{k}^{-1} \mathbb{E}\{\mathbf{g}_{k}^{\rm H} \qg_{k}\}|^{2} =M^2_{\rm B} \lambda_{k}^{2},
\end{equation}
we then have
\begin{equation}\label{denominator1}
\begin{split}
\mathbb{E}|\qz_{2,k}|^2=\underbrace{p_k\tau_{\rm C} M_{\rm B} \lambda_{k}^{2}(\chi_{k}^{-2} \Delta_{k}-M_{\rm B})}_{\Pi_0}.
\end{split}
\end{equation}

Next, we address the right part of the denominator in (\ref{SINR}) as
\begin{equation}\label{Ezik}
\begin{split}
&\mathbb{E}\{\|\qz_{3,k}^{\rm H}-\mathbb{E}\{\qz_{3,k}^{\rm H}\}\|^{2}\} 
=\mathbb{E}\{\|\qz_{3,k}^{\rm H}\|^{2}\}-\|\mathbb{E}\{\qz_{3,k}^{\rm H}\}\|^{2} \\
&=\underbrace{\sum\limits_{i = 1}^4 \mathbb{E}\{\|\qz^{(i)}_{3, k}\|^{2}\}}_{\Pi_1}\underbrace{- \sum\limits_{i = 1}^4 \|\mathbb{E}\{\qz^{(i)}_{3, k}\}\|^{2}}_{\Pi_2} \\
&\underbrace{-2 {\bf {Re}}\{\sum\limits_{i = 1}^4 \sum\limits_{j = i + 1}^4 \mathbb{E}\{\qz^{(i)}_{3, k}\} \mathbb{E}\{(\qz^{(j)}_{3, k})^{\rm H}\}\} }_{\Pi_3}\\
&\underbrace{+2 {\bf {Re}}\{\mathbb{E}\{\sum\limits_{i = 1}^4 \sum\limits_{j=i+1}^4 \qz^{(i)}_{3,k} (\qz^{(j)}_{3,k})^{\rm H}\}\}}_{\Pi_4}.
\end{split}
\end{equation}
In the following, we will calculate $\{\Pi_i\}^4_{i=1}$ respectively.
\subsection{Calculation of $\Pi_1$} 
The expression $\Pi_1$ includes four parts, i.e., $\mathbb{E}\{\|\qz^{(i)}_{3,k}\|^{2}\}, i=1,2,3,4$, which will be addressed one by one.

\subsubsection{ Calculation of $\mathbb{E}\{\|\qz^{(1)}_{3,k}\|^{2}\}$}
\begin{equation}
\begin{array}{l}
\mathbb{E}\{\|\qz^{(1)}_{3,k}\|^{2}\}=p_{k} \mathbb{E}\left\{\left\|{\bar{\qg}}_{k}^{\rm H} \qg_{k} \bm{\nu}_{k}^{\rm H}\right\|^{2}\right\} \\
=p_{k} \mathbb{E}\{\|c_{k} \sum\limits_{i=1}^{K} \sqrt{\frac{p_{i}}{\tau_{{\rm C}}}} \bm{\varphi}_{k}^{\rm H} \bm{\nu}_{i} \qg_{i}^{\rm H} \qg_{k} \bm{\nu}_{k}^{\rm H}  +c_{k} \bm{\varphi}_{k}^{\rm H} \frac{\bm{\Gamma}^{\rm H}}{\sqrt{\tau_{{\rm C}}}} \qg_{k} \bm{\nu}_{k}^{\rm H}\|^{2}\} \\
=p_{k} \Big(\underbrace{\mathbb{E}\{\|c_{k} \sum_{i=1}^{K} \sqrt{\frac{p_{i}}{\tau_{{\rm C}}}} \bm{\varphi}_{k}^{\rm H} \bm{\nu}_{i} \qg_{i}^{\rm H} \qg_{k} \bm{\nu}_{k}^{\rm H}\|^{2}\}}_{A1} \\
+\underbrace{\mathbb{E}\{\|c_{k} \bm{\varphi}_{k}^{\rm H} \frac{\bm{\Gamma}^{\rm H}}{\sqrt{\tau_{{\rm C}}}} \qg_{k} \bm{\nu}_{k}^{\rm H}\|^{2}\}}_{A2}\Big), \\
\end{array}
\end{equation}
where
\begin{equation}
\begin{split}
A_1 & = \underbrace{\mathbb{E}\{\|c_{k} \sqrt{\frac{p_{k}}{\tau_{{\rm C}}}} \bm{\varphi}_{k}^{\rm H} \bm{\nu}_{k} \qg_{k}^{\rm H} \mathbf{g}_{k} \bm{\nu}_{k}^{\rm H}\|^{2}\}}_{A_{11}} \\
& +\underbrace{\mathbb{E}\{\|c_{k} \sum_{i \neq k}^{K} \sqrt{\frac{p_{i}}{\tau_{{\rm C}}}} \boldsymbol{\varphi}_{k}^{\rm H} \bm{\nu}_{i} \mathbf{g}_{i}^{\rm H} \qg_{k} \bm{\nu}_{k}^{\rm H}\|^{2}\}}_{A_{12}},
\end{split}
\end{equation}
where $A_{11}$ and $A_{12}$ are computed as
\begin{equation}\label{A1}
\begin{split}
A_{11} & =c_{k}^{2} \frac{p_{k}}{\tau_{{\rm C}}} \{{\bm \varphi}_{k}^{\rm H} \mathbb{E}\left\{\|\mathbf{g}_{k}\right\|^{4}\} \mathbb{E}\{(\bm{\nu}_{k} \bm{\nu}_{k}^{\rm H})^{2}\} {\bm \varphi}_{k}\} \\
& =M_{\rm B} c_{k}^{2} p_{k}\left(\tau_{{\rm C}}+1\right) \Delta_{k},
\end{split}
\end{equation}
and
\begin{equation}\label{A2}
\begin{split}
A_{12} &= \frac{c_{k}^{2}}{\tau_{\rm C}} \sum_{i \neq k}^{K} \mathbb{E}\left\{p_{i} {\bm \varphi}_{k}^{\rm H} \bm{\nu}_{i} \qg_{i}^{\rm H} \qg_{k} \bm{\nu}_{k}^{\rm H} \bm{\nu}_{k} \qg_{k}^{\rm H} \qg_{i} \bm{\nu}_{i}^{\rm H} {\bm \varphi}_{k}\right\} \\
& = c_{k}^{2} \tau_{\rm C} \sum_{i \neq k}^{K} p_{i} \mathbb{E}\{|\qg_{k}^{\rm H} \qg_{i}|^{2}\} \\
& =M_{\rm B} c_{k}^{2} \tau_{\rm C} \sum_{i \neq k}^{K} p_{i} \Omega_{k i},
\end{split}
\end{equation}
respectively. Therefore, $A_1$ can be expressed as
\begin{equation}
\begin{aligned}
A_1 & =c_{k}^{2} p_{k}\left(\tau_{\rm C}+1\right) M_{\rm B} \Delta_{k} +M_{\rm B} c_{k}^{2} \tau_{\rm C} \sum_{i \neq k}^{K} p_{i} \Omega_{k i}
\end{aligned}
\end{equation}
Besides, $A_2 = M_{\rm B} \tau_{\rm C} c^2_k \chi_k$. Thus, we have $\mathbb{E}\{\|\qz^{(1)}_{3, k}\|^{2}\}=\Pi_{11}$, where
\begin{equation}
\begin{split}
\Pi_{11}&=M_{\rm B} c_{k}^{2} p_{k}^{2} (\tau_{\rm C}+1) \Delta_{k} +M_{\rm B} c_{k}^{2} \tau_{\rm C} \sum_{i \neq k}^{K} p_{i} p_{k} \Omega_{k i} \\
& +M_{\rm B} c_{k}^{2} \tau_{\rm C} p_{k} \chi_{k}.
\end{split}
\end{equation}
\subsubsection{Calculation of $\mathbb{E}\{\|\qz^{(2)}_{3, k}\|^{2}\}$} 
\begin{equation}
\begin{array}{l}
\mathbb{E}\{\|\qz^{(2)}_{3, k}\|^{2}\} 
=\mathbb{E}\{\|\sqrt{p_{i}} \sum\limits_{i \neq k}^{K} \hat{\qg}_{k}^{\rm H} \qg_{i} \mathbf{s}_{i}^{\rm H}\|^{2}\} \\
=\underbrace{\mathbb{E}\{\|c_{k} \sqrt{p_{i}} \sum_{i \neq k}^{K} \sqrt{q_{k} \tau_{\rm C}} \qg_{k}^{\rm H} \qg_{i} \bm{\nu}_{i}^{\rm H}\|^{2}\}}_{B_1} \\
+\underbrace{\mathbb{E}\{\|c_{k} \sqrt{p_{i}} \sum_{i \neq k}^{K} \sum_{j=1}^{K} \sqrt{\frac{p_{j}}{\tau_{\rm C}}} {\bm \varphi}_{k}^{\rm H} \bm{\nu}_{j} \mathbf{g}_{j}^{\rm H} \qg_{i} \bm{\nu}_{i}^{\rm H}\|^{2}\}}_{B_2} \\
+\underbrace{\mathbb{E}\{\|c_{k} \sqrt{p_{i}} \sum_{i \neq k}^{K} \bm{\varphi}_{k}^{\rm H} \frac{\bm{\Gamma}^{\rm H}}{\sqrt{\tau_{\rm C}}} \qg_{i} \bm{\nu}_{i}^{\rm H}\|^{2}\}}_{B_3}
\end{array}
\end{equation}
where
\begin{equation}
\begin{split}
B_1 & =q_{k} c_{k}^{2} \tau_{\rm C}^{2} p_{i} \sum_{i \neq k}^{K} \mathbb{E}\left\{\left|\mathbf{g}_{k}^{\rm H} \mathbf{g}_{i}\right|^{2}\right\} \\
& =M_{\rm B} q_{k} c_{k}^{2} \tau_{\rm C}^{2} \sum_{i \neq k}^{K} p_{i} \Omega_{k i},
\end{split}
\end{equation}
and
\begin{equation}
\begin{split}
B_2 =\underbrace{\mathbb{E}\{\|c_{k} \sqrt{p_{i}} \sum_{i \neq k}^{K} \sqrt{\frac{p_{i}}{\tau_{\rm C}}} \bm{\varphi}_{k}^{\rm H} \bm{\nu}_{i} \qg_{i}^{\rm H} \qg_{i} \bm{\nu}_{i}^{\rm H}\|^{2}\}}_{B_{21}} \\
+\underbrace{\mathbb{E}\{\|c_{k} \sqrt{p_{i}} \sum_{i \neq k}^{K} \sum_{j \neq i}^{K} \sqrt{\frac{p_{j}}{\tau_{\rm C}}} \bm{\varphi}_{k}^{\rm H} \bm{\nu}_{j} \qg_{j}^{\rm H} \mathbf{g}_{i} \bm{\nu}_{i}^{\rm H}\|^{2}\}}_{B_{22}},
\end{split}
\end{equation}
with
\begin{equation}
\begin{aligned}
B_{21} & =\frac{c_{k}^{2}}{\tau_{\rm C}} \mathbb{E}\{\|\sum_{i \neq k}^{K} p_{i} \bm{\varphi}_{k}^{\rm H} \bm{\nu}_{i} \qg_{i}^{\rm H} \qg_{i} \bm{\nu}_{i}^{\rm H}\|^{2}\} \\
& =\frac{c_{k}^{2}}{\tau_{\rm C}} \mathbb{E}\{\sum_{i \neq k}^{K} p_{i}^{2} \bm {\varphi}_{k}^{\rm H}{(\bm{\nu}_{i} \bm{\nu}_{i}^{\rm H})}^{2}\|\mathbf{g}_{i}\|^{4} \bm{\varphi}_{k} \\
& +\sum_{\substack{i \neq k}}^{K} \sum_{\substack{j \neq k \\
j \neq i}}^{K} p_{i} p_{j} \bm {\varphi}_{k}^{\rm H}(\bm{\nu}_{i} \bm{\nu}_{i}^{\rm H})(\bm{\nu}_{j} \bm{\nu}_{j}^{\rm H})\|\qg_{i}\|^{2}\|\qg_{j}\|^{2} \bm {\varphi}_{k}\} \\
& =c_{k}^{2} \sum_{i \neq k}^{K} p_{i}^{2}(\tau_{\rm C}+1) M_{\rm B} \Delta_{i}  +c_{k}^{2} \sum_{i \neq k}^{K} \sum_{\substack{j \neq k \\
j \neq i}}^{K} p_{i} p_{j} M_{\rm B} \Xi_{i j}
\end{aligned}
\end{equation}
and
\begin{equation}
\begin{split}
   B_{22} & = \mathbb{E}\{\|c_{k} \sqrt{p_{i}} \sum_{i \neq k}^{K} \sum_{j \neq i}^{K} \sqrt{\frac{p_{j}}{\tau_{{\rm C}}}} \bm{\varphi}_{k}^{\rm H} \bm{\nu}_{j} \qg_{j}^{\rm H} \qg_{i} \bm{\nu}_{i}^{\rm H}\|^{2}\} \\
     & =M_{\rm B} \tau_{{\rm C}} c_{k}^{2} \sum_{i \neq k}^{K} \sum_{j \neq i}^{K} p_{i} p_{j} \Omega_{i j}.
\end{split}
\end{equation}
Therefore,
\begin{equation}
\begin{split}
B_2 & = M_{\rm B} \tau_{\rm C} c_k^2 \sum_{i \neq k}^K p_i^2 \Delta_i+M_{\rm B} c_k^2 \sum_{i \neq k}^K p_i^2 \Delta_i \\
& +M_{\rm B} c_k^2 \sum_{i \neq k}^K \sum_{\substack{j \neq k \\ j \neq i}}^K p_i p_j \Xi_{i j}  +M_{\rm B} \tau_{\rm C} c_k^2 \sum_{i \neq k}^K \sum_{j \neq i}^K p_i p_j \Omega_{i j},
\end{split}
\end{equation}
and
\begin{equation}
\begin{aligned}
B_3 &=\frac{c_{k}^{2}}{\tau_{{\rm C}}} \mathbb{E}\bigg\{\sqrt{p_{i} p_{j}} \sum_{i \neq k}^{K} \sum_{j \neq k}^{K} \bm{\varphi}_{k}^{\rm H} {\bf \Gamma}^{\rm H} \qg_{i} \bm{\nu}_{i}^{\rm H} \bm{\nu}_{j} \qg_{j}^{\rm H} {\bf \Gamma} {\bm \varphi_{k}}\bigg\} \\
& =M_{\rm B} \tau_{{\rm C}} c_{k}^{2} \sum_{i \neq k}^{K} p_{i} \chi_{i}.
\end{aligned}
\end{equation}
Then, we have $\mathbb{E}\{\|\qz^{(2)}_{3, k}\|^{2}\}=\Pi_{12}$, where
\begin{equation}
\begin{aligned}
\Pi_{12} & =M_{\rm B} q_{k} c_{k}^{2} \tau_{{\rm C}}^{2} \sum_{i \neq k}^{K} p_{i} \Omega_{k i} +M_{\rm B} \tau_{{\rm C}} c_{k}^{2} \sum_{i \neq k}^{K} p_{i}^{2} \Delta_{i} \\
& +M_{\rm B} c_{k}^{2} \sum_{i \neq k}^{K} p_{i}^{2} \Delta_{i}  +M_{\rm B} c_{k}^{2} \sum_{\substack{i \neq k}}^{K} \sum_{\substack{j \neq k \\
j \neq i}}^{K} p_{i} p_{j} \Xi_{i j} \\
& +M_{\rm B} \tau_{{\rm C}} c_{k}^{2} \sum_{i \neq k}^{K} \sum_{j \neq i}^{K} p_{i} p_{j} \Omega_{i j}  +M_{\rm B} \tau_{{\rm C}} c_{k}^{2} \sum_{i \neq k}^{K} p_{i} \chi_{i}.
\end{aligned}
\end{equation}
\subsubsection{ Calculation of  $\mathbb{E}\{\|\qz^{(3)}_{3, k}\|^{2}\}$}
Since
\begin{equation}
\begin{split}
 \mathbb{E}\{\|\qz^{(3)}_{3, k}\|^{2}\}& =\mathbb{E}\{\|\sum_{i=1}^{K} \sqrt{q_{i}} \hat{\qg}_{k}^{\rm H} \bm{\epsilon}_{i} \bm{\varphi}_{i}^{\rm H}\|^{2}\},
\end{split}
\end{equation}   
we have $\mathbb{E}\{\|\qz^{(3)}_{3, k}\|^{2}\}=\Pi_{13}$, where
\begin{equation}
    \Pi_{13}=M_{\rm B} \tau_{{\rm C}} \sum_{i=1}^{K} q_{i} \lambda_{k}(\chi_{i}-\lambda_{i}).
\end{equation}
\subsubsection{Calculation of $\mathbb{E}\{\|\mathbf{z}^{(4)}_{3, k}\|^{2}\}$}
\begin{equation}
\begin{split}
& \mathbb{E}\{\|\mathbf{z}^{(4)}_{3, k}\|^{2}\} 
 =\underbrace{\mathbb{E}\{\|c_{k} \sqrt{q_{k} \tau_{\rm C}} \qg_{k}^{\rm H} \bm{\Gamma}\|^{2}\}}_{C_1} \\
& +\underbrace{\mathbb{E}\{\|c_{k} \sum\limits_{i=1}^{K} \sqrt{\frac{p_{i}}{\tau_{\rm C}}} \bm{\varphi}_{k}^{\rm H} \bm{\nu}_{i} \mathbf{g}_{i}^{\rm H} \bm{\Gamma}\|^{2}\}}_{C_2} +\underbrace{\mathbb{E}\{\|c_{k} \bm{\varphi}_{k}^{\rm H} \frac{\bm{\Gamma}^{\rm H} \bm{\Gamma}}{\sqrt{\tau_{\rm C}}}\|^{2}\}}_{C_3},
\end{split}
\end{equation}
where
\begin{equation}
\begin{split}
C_1 & = \mathbb{E}\{\|c_k \sqrt{q_k \tau_{\rm C}} \qg_k^{\rm H} \bm{\Gamma}\|^2\}\\
& = c_k^2 q_k \tau_{\rm C} \mathbb{E}\{\qg_k^{\rm H} \bm{\Gamma} \bm{\Gamma}^{\rm H} \qg_k\}=M_{\rm B} c_k^2 q_k \tau_{\rm C}^2 \chi_k,
\end{split}
\end{equation}
and
\begin{equation}
\begin{split}
C_2& = \mathbb{E}\{\|c_k \sum_{i=1}^K \sqrt{\frac{p_i}{\tau_{\rm C}}} \bm{\varphi}_k^{\rm H} \bm{\nu}_i \qg_i^{\rm H} \bm{\Gamma}\|^2\} \\
& =\frac{c_k^2}{\tau_{\rm C}} \sum_{i=1}^K p_i \mathbb{E}\{\bm{\varphi}_k^{\rm H} \bm{\nu}_i \qg_i^{\rm H} \bm{\Gamma} \bm{\Gamma}^{\rm H} \qg_i \bm{\nu}_i^{\rm H} \bm{\varphi}_k\} \\
& =M_{\rm B} c_k^2 \tau_{\rm C} \sum_{i=1}^K p_i \chi_i,
\end{split}
\end{equation}
and
\begin{equation}
\begin{split}
C_3 & = \mathbb{E}\{\|c_k \bm{\varphi}_k^{\rm H} \frac{\bm{\Gamma}^{\rm H} \bm{\Gamma}}{\sqrt{\tau_{\rm C}}}\|^2\}\\
 & = \frac{c_k^2}{\tau_{\rm C}} \mathbb{E}\{\|\bm{\varphi}_k^{\rm H} \bm{\Gamma}^{\rm H} \bm{\Gamma} \bm{\Gamma}^{\rm H}\bm{\Gamma} \bm{\varphi}_k\|^2\}\\
 & = c_k^2 M_{\rm B}(M_{\rm B}+\tau_{\rm C}).
\end{split}
\end{equation}
Thus, we have $\mathbb{E}\{\|\mathbf{z}^{(4)}_{3,k}\|^{2}\}=\Pi_{14}$, where
\begin{equation}
\begin{aligned}
\Pi_{14}
& =M_{\rm B} \lambda_k \tau_{\rm C}+c_k^2 M^2_{\rm B}.
\end{aligned}
\end{equation}
Finally, we have $\Pi_1=\sum_{i=1}^4\Pi_{1i}$.

\subsection{Calculation of $\Pi_2$, $\Pi_3$, and $\Pi_4$ }
\subsubsection{Calculation of $\Pi_2$}
The expression $\Pi_2$ also has four parts, which are computed as
\begin{equation}
\begin{split}
\|\mathbb{E}\{\qz^{(1)}_{3, k}\}\|^2  
 & =M^2_{\rm B} p_k^2 c_k^2 \chi_k^2,\\
 \|\mathbb{E}\{\qz^{(2)}_{3, k}\}\|^2
& =M^2_{\rm B} c_k^2(\sum_{i \neq k}^K p_i \chi_i)^2,\\
\|\mathbb{E}\{\qz^{(3)}_{3,k}\}\|^2& = 0,\\
\|\mathbb{E}\{\qz^{(4)}_{3,k}\}\|^2 &= M^2_{\rm B}c^2_k,
\end{split}
\end{equation} respectively. Thus, we have
\begin{equation}
   \Pi_2=-[M^2_{\rm B} p_k^2 c_k^2 \chi_k^2+M^2_{\rm B} c_k^2(\sum_{i \neq k}^K p_i \chi_i)^2+M^2_{\rm B}c^2_k]. 
\end{equation}

\subsubsection{Calculation of $\Pi_3$}
Since
\begin{equation}
\begin{aligned}
& {\bf {Re}}\Big\{\sum_{i=1}^4 \sum_{j=i+1}^4 \mathbb{E}\{\qz^{(i)}_{3, k}\} \mathbb{E}\{(\qz^{(j)}_{3, k})^{\rm H}\}\Big\} \\
& ={\bf {Re}}\Big\{\mathbb{E}\{\qz^{(1)}_{3, k}\} \mathbb{E}\{(\qz^{(2)}_{3, k})^{\rm H}\} +\mathbb{E}\{\qz^{(1)}_{3, k}\} \mathbb{E}\{(\qz^{(4)}_{3, k})^{\rm H}\}\\
&+\mathbb{E}\{\qz^{(2)}_{3, k}\} \mathbb{E}\{(\qz^{(4)}_{3, k})^{\rm H}\}\Big\},
\end{aligned}
\end{equation}
then we have
\begin{equation}
\Pi_3= -2M^2_{\rm B} c_k^2 \Big(p_k \chi_k \sum_{i \neq k}^K p_i \chi_i  + p_k \chi_k+  \sum_{i \neq k}^K p_i \chi_i\Big).
\end{equation}

\subsubsection{Calculation of $\Pi_4$}
Similarly,
\begin{equation}
\begin{aligned}
&{\bf {Re}}\Big\{\mathbb{E}\{\sum_{i=1}^4 \sum_{j=i+1}^4 \qz^{(i)}_{3, k} (\qz^{(j)}_{3, k})^{\rm H}\}\Big\} \\
& ={\bf {Re}}\Big\{\mathbb{E}\{\qz^{(1)}_{3, k} (\qz^{(2)}_{3,k})^{\rm H}\}+\mathbb{E}\{\qz^{(1)}_{3, k} (\qz^{(4)}_{3, k})^{\rm H}\}+\mathbb{E}\{\qz^{(2)}_{3,k} (\qz^{(4)}_{3, k})^{\rm H}\}\Big\},
\end{aligned}
\end{equation}
then we have
\begin{equation}
\Pi_4=2M_{\rm B} c_k^2 \Big(p_k \sum_{i \neq k}^K p_i \Xi_{i k}+ M_{\rm B} p_k \chi_k  + M_{\rm B} \sum_{i \neq k}^K p_i \chi_i\Big),
\end{equation}
Therefore, we have
\begin{equation}\label{closed form of noise}
\mathbb{E}\{\|\qz_{3,k}-\mathbb{E}\{\qz_{3,k}\}\|^2\} = \sum_{i=1}^4\Pi_i.
\end{equation}
By substituting \eqref{numerator}, \eqref{denominator1}, and \eqref{closed form of noise} into \eqref{SINR}, then the proof is completed.
\end{appendices}


\begin{thebibliography}{10}
\providecommand{\url}[1]{#1}
\csname url@samestyle\endcsname
\providecommand{\newblock}{\relax}
\providecommand{\bibinfo}[2]{#2}
\providecommand{\BIBentrySTDinterwordspacing}{\spaceskip=0pt\relax}
\providecommand{\BIBentryALTinterwordstretchfactor}{4}
\providecommand{\BIBentryALTinterwordspacing}{\spaceskip=\fontdimen2\font plus
\BIBentryALTinterwordstretchfactor\fontdimen3\font minus
  \fontdimen4\font\relax}
\providecommand{\BIBforeignlanguage}[2]{{%
\expandafter\ifx\csname l@#1\endcsname\relax
\typeout{** WARNING: IEEEtran.bst: No hyphenation pattern has been}%
\typeout{** loaded for the language `#1'. Using the pattern for}%
\typeout{** the default language instead.}%
\else
\language=\csname l@#1\endcsname
\fi
#2}}
\providecommand{\BIBdecl}{\relax}
\BIBdecl

\bibitem{tang2021wireless}
W.~Tang, M.~Z. Chen, X.~Chen, J.~Y. Dai, Y.~Han, M.~Di~Renzo, Y.~Zeng, S.~Jin,
  Q.~Cheng, and T.~J. Cui, ``Wireless communications with reconfigurable
  intelligent surface: {Path} loss modeling and experimental measurement,''
  \emph{IEEE Trans. Wireless Commun.}, vol.~20, no.~1, pp. 421--439, Jan. 2021.

\bibitem{bjornson2022reconfigurable}
E.~Bj{\"o}rnson, H.~Wymeersch, B.~Matthiesen, P.~Popovski, L.~Sanguinetti, and
  E.~De~Carvalho, ``{Reconfigurable intelligent surfaces: A signal processing
  perspective with wireless applications},'' \emph{IEEE Signal Process.
  Magazine}, vol.~39, no.~2, pp. 135--158, Mar. 2022.

\bibitem{li2024ergodic}
Q.~Li, M.~El-Hajjar, and L.~Hanzo, ``Ergodic spectral efficiency analysis of
  intelligent omni-surface aided systems suffering from imperfect {CSI} and
  hardware impairments,'' \emph{IEEE Transactions on Communications}, vol.~72,
  no.~8, pp. 5073--5086, 2024.

\bibitem{wang2020channel}
Z.~Wang, L.~Liu, and S.~Cui, ``{Channel estimation for intelligent reflecting
  surface assisted multiuser communications: Framework, algorithms, and
  analysis},'' \emph{IEEE Trans. Wireless Commun.}, vol.~19, no.~10, pp.
  6607--6620, 2020.

\bibitem{hu2021two}
C.~Hu, L.~Dai, S.~Han, and X.~Wang, ``Two-timescale channel estimation for
  reconfigurable intelligent surface aided wireless communications,''
  \emph{IEEE Trans. Commun.}, vol.~69, no.~11, pp. 7736--7747, 2021.

\bibitem{you2020channel}
C.~You, B.~Zheng, and R.~Zhang, ``Channel estimation and passive beamforming
  for intelligent reflecting surface: Discrete phase shift and progressive
  refinement,'' \emph{IEEE J. Sel. Areas Commun.}, vol.~38, no.~11, pp.
  2604--2620, Nov. 2020.

\bibitem{hoeher1999channel}
P.~Hoeher and F.~Tufvesson, ``Channel estimation with superimposed pilot
  sequence,'' in \emph{Proc. Global Telecommunications Conference (GLOBECOM)},
  vol.~4, Rio de Janeiro, Brazil, Dec. 1999, pp. 2162--2166.

\bibitem{tong2004pilot}
L.~Tong, B.~Sadler, and M.~Dong, ``Pilot-assisted wireless transmissions:
  {General} model, design criteria, and signal processing,'' \emph{IEEE Signal
  Process. Mag.}, vol.~21, no.~6, pp. 12--25, Nov. 2004.

\bibitem{asyhari2017orthogonal}
A.~T. Asyhari and S.~ten Brink, ``Orthogonal or superimposed pilots? {A}
  rate-efficient channel estimation strategy for stationary {MIMO} fading
  channels,'' \emph{IEEE Trans. Wireless Commun.}, vol.~16, no.~5, pp.
  2776--2789, May 2017.

\bibitem{verenzuela2018spectral}
D.~Verenzuela, E.~Bj{\"o}rnson, and L.~Sanguinetti, ``Spectral and energy
  efficiency of superimposed pilots in uplink massive {MIMO},'' \emph{IEEE
  Trans. Wireless Commun.}, vol.~17, no.~11, pp. 7099--7115, Nov. 2018.

\bibitem{jing2018superimposed}
X.~Jing, M.~Li, H.~Liu, S.~Li, and G.~Pan, ``Superimposed pilot optimization
  design and channel estimation for multiuser massive {MIMO} systems,''
  \emph{IEEE Trans. Veh. Technol.}, vol.~67, no.~12, pp. 11\,818--11\,832, Dec.
  2018.

\bibitem{zhang2021superimposed}
Y.~Zhang, X.~Qiao, L.~Yang, and H.~Zhu, ``Superimposed pilots are beneficial
  for mitigating pilot contamination in cell-free massive {MIMO},'' \emph{IEEE
  Commun. Lett.}, vol.~25, no.~1, pp. 279--283, Jan. 2021.

\bibitem{weiss2003accuracy}
A.~J. Weiss, ``{On the accuracy of a cellular location system based on RSS
  measurements},'' \emph{IEEE Trans. Veh. Technol.}, vol.~52, no.~6, pp.
  1508--1518, 2003.

\bibitem{keykhosravi2021siso}
K.~Keykhosravi, M.~F. Keskin, G.~Seco-Granados, and H.~Wymeersch, ``{SISO
  RIS-enabled joint 3D downlink localization and synchronization},'' in
  \emph{Proc. IEEE Int. Conf. Commun. (ICC)}, Montreal, Canada, June 2021, pp.
  1--6.

\bibitem{hu2023irs}
X.~Hu, C.~Liu, M.~Peng, and C.~Zhong, ``{IRS-based integrated location sensing
  and communication for mmWave SIMO systems},'' \emph{IEEE Trans. Wireless
  Commun.}, vol.~22, no.~6, pp. 4132--4145, 2023.

\bibitem{hu2018beyond}
S.~Hu, F.~Rusek, and O.~Edfors, ``Beyond massive mimo: The potential of
  positioning with large intelligent surfaces,'' \emph{IEEE Trans. Signal
  Process.}, vol.~66, no.~7, pp. 1761--1774, 2018.

\bibitem{wymeersch2020beyond}
H.~Wymeersch and B.~Denis, ``Beyond 5g wireless localization with
  reconfigurable intelligent surfaces,'' in \emph{Proc. IEEE Int. Conf. Commun.
  (ICC)}, 2020, pp. 1--6.

\bibitem{he2020large}
J.~He, H.~Wymeersch, L.~Kong, O.~Silv{\'e}n, and M.~Juntti, ``Large intelligent
  surface for positioning in millimeter wave {MIMO} systems,'' in \emph{Proc.
  IEEE Veh. Technol. Conf. (VTC-Spring)}, 2020, pp. 1--5.

\bibitem{chen2024joint}
Z.~Chen, M.-M. Zhao, M.~Li, F.~Xu, Q.~Wu, and M.-J. Zhao, ``{Joint location
  sensing and channel estimation for IRS-aided mmWave ISAC systems},''
  \emph{IEEE Trans. Wireless Commun.}, 2024.

\bibitem{fascista2022ris}
A.~Fascista, M.~F. Keskin, A.~Coluccia, H.~Wymeersch, and G.~Seco-Granados,
  ``{RIS-aided joint localization and synchronization with a single-antenna
  receiver: Beamforming design and low-complexity estimation},'' \emph{IEEE J.
  Sel. Topics Signal Process.}, vol.~16, no.~5, pp. 1141--1156, 2022.

\bibitem{jiang2022optimization}
F.~Jiang, A.~Abrardo, K.~Keykhosravi, H.~Wymeersch, D.~Dardari, and
  M.~Di~Renzo, ``Two-timescale transmission design and {RIS} optimization for
  integrated localization and communications,'' \emph{IEEE Trans. Wireless
  Commun.}, vol.~22, no.~12, pp. 8587--8602, Apr. 2023.

\bibitem{peng2024joint}
X.~Peng, X.~Hu, X.~Gan, and C.~Zhong, ``Joint location sensing and demodulation
  for {IRS}-assisted {ISAC mmWave MIMO} systems,'' \emph{IEEE Trans. Commun.},
  vol.~72, no.~4, pp. 2470--2484, 2024.

\bibitem{9903389}
M.~Cui, Z.~Wu, Y.~Lu, X.~Wei, and L.~Dai, ``Near-field {MIMO} communications
  for {6G: Fundamentals}, challenges, potentials, and future directions,''
  \emph{IEEE Communications Magazine}, vol.~61, no.~1, pp. 40--46, 2023.

\bibitem{li2023achievable}
Q.~Li, M.~El-Hajjar, Y.~Sun, I.~Hemadeh, A.~Shojaeifard, Y.~Liu, and L.~Hanzo,
  ``{Achievable rate analysis of the STAR-RIS-aided NOMA uplink in the face of
  imperfect CSI and hardware impairments},'' \emph{IEEE Transactions on
  Communications}, vol.~71, no.~10, pp. 6100--6114, 2023.

\bibitem{10025392}
Q.~Li, M.~El-Hajjar, I.~Hemadeh, D.~Jagyasi, A.~Shojaeifard, and L.~Hanzo,
  ``Performance analysis of active ris-aided systems in the face of imperfect
  {CSI} and phase shift noise,'' \emph{IEEE Transactions on Vehicular
  Technology}, vol.~72, no.~6, pp. 8140--8145, 2023.

\bibitem{Melanie1998An}
M.~Mitchell, \emph{An Introduction to Genetic Algorithms}.\hskip 1em plus 0.5em
  minus 0.4em\relax Cambridge, MA, USA: MIT Press, 1998.

\bibitem{Zhi2022Power}
K.~Zhi, C.~Pan, H.~Ren, and K.~Wang, ``Power scaling law analysis and phase
  shift optimization of {RIS-Aided} massive {MIMO} systems with statistical
  {CSI},'' \emph{IEEE Trans. Commun.}, vol.~70, no.~5, pp. 3558--3574, May
  2022.

\bibitem{Peng2021Analysis}
Z.~Peng, T.~Li, C.~Pan, H.~Ren, W.~Xu, and M.~D. Renzo, ``Analysis and
  optimization for {RIS-Aided} multi-pair communications relying on statistical
  {CSI},'' \emph{IEEE Trans. Veh. Technol.}, vol.~70, no.~4, pp. 3897--3901,
  Apr. 2021.

\bibitem{Dai2022Statistical}
J.~Dai, F.~Zhu, C.~Pan, H.~Ren, and K.~Wang, ``Statistical {CSI}-based
  transmission design for reconfigurable intelligent surface-aided massive
  {MIMO} systems with hardware impairments,'' \emph{IEEE Wireless Commun.
  Lett.}, vol.~11, no.~1, pp. 38--42, Jan. 2022.

\end{thebibliography}
\end{document}